\pdfoutput=1
\documentclass[cernpreprint, subfigure, UKenglish, texmf, orcidlogo]{atlasdoc}
\usepackage{atlaspackage}
\usepackage{arydshln}
\usepackage{multirow}
\usepackage{graphicx}
\usepackage{graphbox}
\usepackage{seqsplit}
\usepackage{amsfonts}
\usepackage{makecell}
\usepackage{atlasbiblatex}

\usepackage[jetetmiss,BSM]{atlasphysics}
\usepackage{verbatim}
\graphicspath{{logos/}{figures/}}

\usepackage{ANA-HMBS-2024-68-PAPER-defs}
\AtlasTitle{Search for long-lived charged particles using large specific ionisation loss and time of flight in 140 $fb^{-1}$ of $pp$ collisions at $\sqrt{s} = 13$ TeV with the ATLAS detector}

\AtlasAbstract{%
This paper presents a search for massive, charged, long-lived
particles with the ATLAS detector at the Large Hadron Collider using an
integrated luminosity of 140 $fb^{-1}$ of proton-proton
collisions at $\sqrt{s}=13$ TeV. These particles are expected
to move significantly slower than the speed of light. In this paper,
two signal regions provide complementary sensitivity. In one region,
events are selected with at least one charged-particle track with high transverse momentum, large specific
ionisation measured in the pixel detector, and time of flight
to the hadronic calorimeter inconsistent with the speed of light. In
the other region, events are selected with at least two tracks of opposite charge which both have a high transverse momentum and an anomalously large specific ionisation. The search is sensitive to particles with lifetimes greater than about 3 ns with masses ranging from 200 GeV to 3 TeV. The results are interpreted to set constraints on the supersymmetric pair production of long-lived R-hadrons, charginos and staus, with mass limits extending beyond those from previous searches in broad ranges of lifetime.
}

\AtlasRefCode{HMBS-2024-68}

\PreprintIdNumber{CERN-EP-2025-008}

\AtlasJournal{JHEP}
\AtlasJournalRef{\JHEP 07 (2025) 140}
\AtlasDOI{10.1007/JHEP07(2025)140}

\hypersetup{pdftitle={ATLAS document},pdfauthor={The ATLAS Collaboration}}

\begin{document}

\maketitle

\setcounter{tocdepth}{3}
\tableofcontents

\section{Introduction}
\label{sec:intro}

\FloatBarrier
A wide range of physics models that extend the Standard Model (SM) predict the existence of new, massive, charged long-lived particles (LLPs). These particles appear in certain dark matter models~\cite{Acaro_lu_2023,asadi2023duet} and in proposed solutions to the gauge hierarchy problem, including supersymmetric (SUSY) models~\cite{Golfand:1971iw,Volkov:1973ix,Wess:1974tw,Wess:1974jb,Ferrara:1974pu,Salam:1974ig} that either violate or conserve $R$-parity. \footnote{$R$-parity is a quantum number defined as $(-1)^{3(B-L)+2S}$ where $S$ is the particle spin and $L$ and $B$ are, respectively, its lepton and baryon number.} Particles can acquire macroscopic lifetimes in models of new physics via the same mechanisms that generate long-lived SM particles, with lifetimes depending on the mass hierarchies between new particles and/or the size of a new coupling.

\begin{figure}[tbp]
\begin{center}
\subfigure[]{
\label{fig:diagram_gluino}
\includegraphics[align=c,height=35mm]{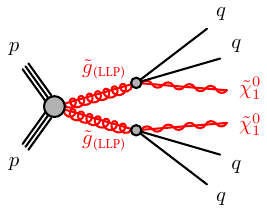}
}
\subfigure[]{
\label{fig:diagram_chargino}
\includegraphics[align=c,height=35mm]{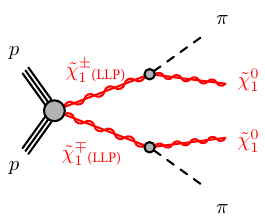}
}
\subfigure[]{
\label{fig:diagram_slepton}
\includegraphics[align=c,height=35mm]{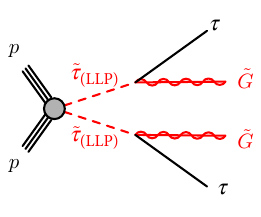}
}
\caption{Representative production diagrams for (a) pair-produced gluinos which form $R$-hadrons decaying into neutralinos, (b)
pair-produced charginos decaying into neutralinos, and (c) pair-produced staus decaying into gravitinos. The anti-particle labels are suppressed for simplicity.}
\label{fig:diagrams}
\end{center}
\end{figure}

A search is presented for particles that are massive, long-lived, and charged using 140 fb$^{-1}$ of proton--proton collision data from the ATLAS experiment at the Large Hadron Collider (LHC)~\cite{Evans:2008zzb}. This analysis looks for a direct interaction of the LLPs with the ATLAS detector using the measurement of the ionisation energy loss (\dEdx) in the pixel detector and the time of flight (ToF) measured by the hadronic calorimeter. This search is, to first order, independent of the LLP decay mode and is therefore sensitive to many different models of new physics. The analysis is optimised for and interpreted in the context of several different long-lived, pair-produced, supersymmetric particles. Long-lived gluinos appear in models including mini-split SUSY~\cite{Giudice:2004tc,ArkaniHamed:2004fb}; once produced, the gluinos hadronise with Standard Model quarks to produce $R$-hadrons~\cite{Fairbairn:2006gg}. Long-lived charginos are motivated by anomaly mediated supersymmetry-breaking (AMSB) models~\cite{Giudice:1998xp,Randall:1998uk}. Long-lived staus emerge in both co-annihilation dark matter models~\cite{Ellis:1998kh,Jittoh:2005pq,Kaneko:2008re} and in gauge-mediated supersymmetry-breaking (GMSB) models~\cite{Dine:1981gu,AlvarezGaume:1981wy,Nappi:1982hm}. Representative production and decay diagrams of the processes targeted by this search are shown in Figure~\ref{fig:diagrams}.

This search is designed to extend the reach of a previous paper~\cite{SUSY-2018-42}, in which the primary signal selection requirement was an isolated, high-momentum track with large \dEdx. Two complementary search strategies extend the sensitivity with respect to the previous result for different signal topologies.

One analysis region ($\beta$-search) extends the sensitivity to higher masses for models with one or more charged, heavy LLPs by requiring that the tracks that pass signal candidate selections have both a large \dEdx and a ToF measurement consistent with a slow-moving particle. The additional ToF requirement reduces the contribution of background processes and therefore improves the sensitivity to models in which a large \dEdx in the pixel detector arises from a heavy, slow-moving particle, with a lifetime greater than about 10 ns. This analysis specifically targets heavy LLPs, including long-lived charginos and $R$-hadrons. By explicitly requiring $\beta\lesssim0.8$, this channel is insensitive to models that predict large \dEdx from relativistic LLPs with an electric charge greater than one~\cite{Giudice:2022bpq}. ATLAS reported on a search for long-lived multi-charged particles in Ref.~\cite{EXOT-2018-54}.

The second analysis region (di-track search) requires two signal tracks which both have significant \dEdx. As the requirement of a second signal track significantly reduces backgrounds, other selection requirements can be relaxed to enhance the signal significance, in particular for relatively light LLPs, which have a more modest ionisation signature. As this region is only sensitive to signatures with two charged LLPs, it does not add sensitivity for $R$-hadrons (which hadronise into a mix of charged and neutral states), nor for charginos produced in association with a neutralino. However, the large reduction in background and enhanced acceptance for low-mass LLPs yields significant sensitivity gains for pair-produced sleptons with lifetimes greater than about 3~ns.

For both signal regions, the main observable used is the mass of the particle associated with the selected track(s). The candidate LLP mass is calculated directly via the relation $m\equiv p/\beta\gamma$ with two independent determinations of $\beta\gamma$, and $p$ measured using the track curvature in the central magnetic field of ATLAS. In both regions, $\bgdEdx$ is extracted from a parameterisation of the Bethe-Bloch relationship between $\beta\gamma$ and \dEdx~\cite{SUSY-2016-31}. The di-track search obtains two independent $\bgdEdx$ measurements per event, one per track, while the $\beta$-search obtains both $\bgdEdx$ and $\bgToF$ for the same track. The $\bgToF$ is calculated from the ToF measured by a cluster of cells in the ATLAS calorimeter crossed by the candidate track and their distance from the proton--proton collision. Compatibility between the two mass measurements is finally required to maximise sensitivity.

This analysis uses the full Run~2 data sample and is an update of several previous searches performed by the ATLAS experiment in both Run~1 and Run~2~\cite{SUSY-2011-03, SUSY-2014-09, SUSY-2016-03, SUSY-2016-31, SUSY-2018-42}. The CMS experiment has also used a combination of \dEdx and ToF in previous searches~\cite{SUSY-2016-32,SUSY-2016-02,CMS-EXO-15-010, CMSLLP2024}. ATLAS observed a 3.3 (3.6) global (local) Z significance excess at 1.4~\TeV\ using only the pixel \dEdx, in 140~fb$^{-1}$ of Run 2 collisions~\cite{SUSY-2018-42}. As reported in Ref.~\cite{SUSY-2018-42}, preliminary, uncalibrated ToF measurements of the calorimeter and muon systems for the tracks in the excess were not compatible with the hypothesis of slow massive particles. The search presented in this paper follows up on this excess by calibrating the ToF measurement in the calorimeter and by designing a signal region with enhanced sensitivity to heavy, charged, and slow LLPs with unit charge, while additionally extending the sensitivity to sleptons with moderate lifetimes with a new di-track region.

\FloatBarrier


%

%
\section{ATLAS detector}
\label{sec:detector}

The ATLAS detector~\cite{PERF-2007-01} is a general-purpose detector with a forward--backward-symmetric cylindrical layout\footnote{ATLAS uses a right-handed coordinate system with its origin at the nominal interaction point in the centre of the detector and the $z$-axis coinciding with the axis of the beam pipe. The $x$-axis points from the interaction point to the centre of the LHC ring, and the $y$-axis points upward. Cylindrical coordinates ($r$, $\phi$) are used in the transverse plane, $\phi$ being the azimuthal angle around the $z$-axis. The pseudorapidity is defined in terms of the polar angle $\theta$ as $\eta$ = $-$ln tan($\theta$/2).} covering nearly $4\pi$ in solid angle.
It consists of an inner detector (ID) tracking system which measures the trajectories of charged particles, surrounded by a 2~T solenoid, followed by calorimeters which measure the energy of particles that interact electromagnetically or hadronically, and a muon spectrometer (MS) inside toroidal magnets which provide additional tracking for muons. The detector is hermetic within its $\eta$ acceptance and can therefore measure the missing transverse momentum ($\vec{p}_\mathrm{T}^{~\mathrm{miss}}$, with magnitude $\MET$) associated with each event.
A two-level trigger system is used to select events~\cite{TRIG-2016-01}. The first-level trigger is hardware-based and uses a subset of detector information to accept  events, produced by LHC at 40 MHz bunch crossing, at a rate below 100 kHz, which is the maximum detector readout rate. This is followed by a software-based high-level trigger, which runs calibration and prompt reconstruction algorithms, reducing the event recording rate to about 1~kHz. The events are eventually processed offline and reconstructed by making use of a software suite~\cite{SOFT-2022-02}, which also provides tools for data simulation, analysis, detector operations, trigger and data acquisition.

Two detectors, the pixel and calorimeter subsystems, are used to measure the $\beta\gamma$ of charged particles and are therefore described in more detail.
The pixel detector~\cite{ATLAS-TDR-11,ATLAS-TDR-19,PIX-2018-001} covers the innermost region of the ID and provides, on average, four precision measurements for each track in the region $|\eta| <2.5$ at radial distances of 3.4~cm to 13~cm from the LHC beam line.
These measurements determine both the track parameters and the charge released in each pixel. The charge is measured by digitising the time interval with the signal above a preset threshold (time-over-threshold or \ToT), which is approximately proportional to the ionisation charge~\cite{BaBar_ToT}. Compared with the other layers that can measure a charge corresponding to ~10 minimum ionising particles (MIPs), the innermost pixel layer, called Insertable B-Layer (IBL)~\cite{ATLAS-TDR-19,PIX-2018-001} provides charge measurements with lower resolution and dynamic range. If the charge released in a pixel exceeds the IBL dynamic range (which is set at approximately two MIPs) an overflow bit is set.
The charge released by a track crossing a layer of the pixel detector is rarely contained within just one pixel; neighbouring pixels registering hits are joined together using a connected component analysis~\cite{PERF-2012-05} to form clusters. The charge of a cluster is calculated by summing the charges of all pixels belonging to the cluster.
The \dEdx measurement assigned to each track is then calculated by averaging the ionisation measurements (charge collected in the cluster per unit track length in the sensor) of its individual clusters. To reduce the effect of the tails of the individual ionisation measurements on this distribution, a truncated average (\dEdxtrunc) is evaluated after removing the highest \dEdx cluster, or the two highest \dEdx clusters in the rare case of more than four pixel clusters on a track. Clusters including pixels at the sensor edges and clusters with overflow ionisation in the IBL are excluded from the \dEdxtrunc calculation, as full charge detection is not guaranteed for these clusters.
A track is considered for this analysis if the \dEdxtrunc is calculated using at least two clusters after removal of those meeting the criteria defined above. The average number of clusters used for the \dEdxtrunc calculation is approximately $2.7$ per track.
The \dEdxtrunc is then corrected for variations of the pixel detector conditions during the data-taking period (e.g.\ charge losses due to radiation damage) and for the residual $\eta$-dependence, as described in Section~\ref{sec:analysis}. The output is the variable used in the signal selection for the search, and is further referred to simply as \dEdx. Like the \textit{restricted energy loss}~\cite{Zyla:2020zbs}, this variable rejects high ionisation deposits.

Neither of these variables show a logarithmic rise at high values of $\beta\gamma$ nor sensitivity to radiative effects, which is expected based on the performance of the restricted energy loss in thin silicon sensors~\cite{Zyla:2020zbs} and confirmed for the specific calculation of \dEdx in the ATLAS pixel detector with dedicated samples of electrons and muons from $Z\rightarrow e e$ and $Z\rightarrow\mu\mu$ events selected in data. The $\beta\gamma$ of a particle can be calculated from the \dEdx of its track using the Bethe--Bloch formula. A meaningful $\beta\gamma$ value can only be estimated in the range of $0.3\lesssim\beta\gamma\lesssim 0.9$ using the pixel detector. The lower limit is a consequence of the \ToT dynamic range, while the upper limit is due to the proximity of the MIP regime which begins at $\beta\gamma\approx$~3 and where \dEdx  becomes quasi-independent of $\beta\gamma$.

A silicon microstrip track detector (SCT)~\cite{SCTD-2019-01} surrounds the pixel detector and contributes to the definition of an accepted track, which must reach a 45 cm radial distance from the colliding beams.

\begin{figure}[tbp]
\begin{center}
\includegraphics[width=0.85\textwidth]{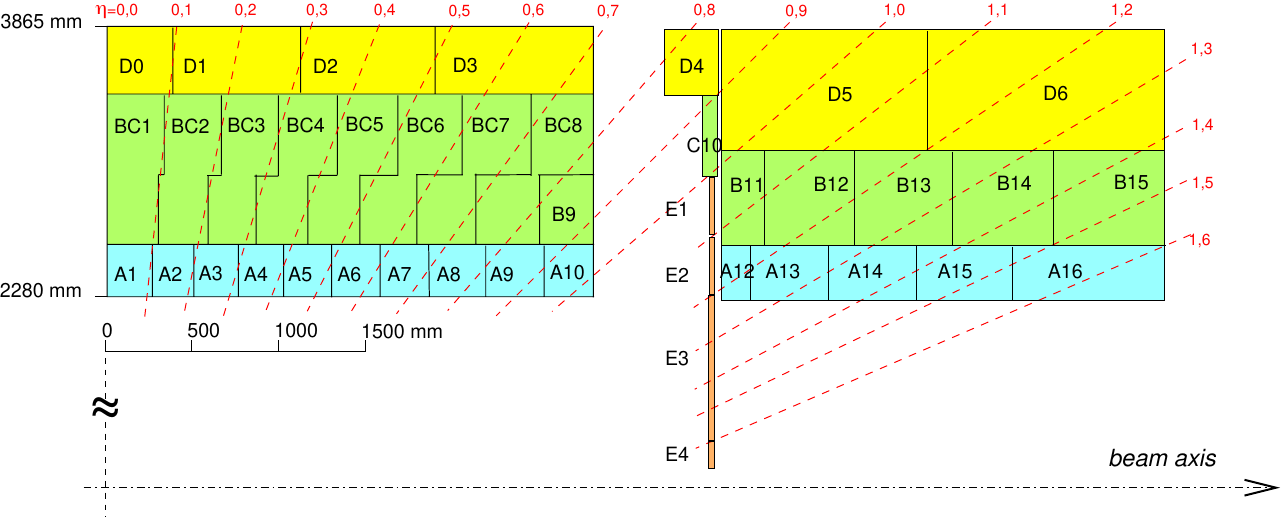}
\caption{Schematic showing the TileCal cell layout and $|\eta|$ acceptance. The red dashed lines indicate where tracks from the origin with a given pseudorapidity will cross the calorimeter, the calorimeter cells A, (B, BC) and D belong to layers at increasing radius and all contribute to the ToF measurement.  The special E-cells are not used in the analysis because their time resolution is poor. The calorimeter response is worse in the region $0.8<|\eta|<$1.0 (transition region between the barrel and the extended barrel).}
\label{fig:TileCal}
\end{center}
\end{figure}

The ATLAS calorimeter system is composed of two parts optimised to measure the energy of the particles interacting electromagnetically or hadronically. As the target LLPs are not expected to shower in the electromagnetic calorimeter and the timing resolution of the electromagnetic calorimeter decreases for small energy deposits, the ToF measurement is done with the central hadronic calorimeter (TileCal)~\cite{TCAL-2017-01} and uses, as a reference, the beam crossing time signal provided by the LHC. The TileCal is a barrel-shaped sampling device (made of steel plates acting as absorber and scintillator tiles as active medium) extending from a radius of 228~cm to 386.5~cm and covering the range $|\eta|<$1.6 as shown in Figure~\ref{fig:TileCal}.

Wavelength-shifting fibres collect the light from scintillators and carry it to the photomultiplier tubes (PMTs). The analogue signals from the PMTs are amplified, shaped and digitised by sampling the signal every 25 ns and stored on detector until a trigger decision is received. The front-end electronics read out the signals produced by approximately 5000 cells organised into three radial layers. The tile calorimeter cells have a good signal time resolution (better than 1.0~ns (3.7~ns) when more than 5~\GeV\ (500~\MeV) are released in one cell) and can therefore determine $\betacalo$ through a ToF measurement. Each calorimeter cell along the particle track contributes to the $\betacalo$ measurement through a weighted average. This takes into account both the cell distance from the interaction point and the energy released in each cell, as the cell time resolution depends on the deposited energy~\cite{TileTimeNoteRun2}.


%

%

\section{Data and Monte Carlo samples}
\label{subsec:samples}


The analysis is conducted with 140 $\text{fb}^{-1}$ of $pp$ collision data that satisfy the ATLAS data quality requirements~\cite{DAPR-2018-01}. The data sample was taken during Run~2 of the LHC from 2015 to 2018, at a centre-of-mass energy of 13~\TeV. The average number of collisions per bunch-crossing (\pileup, $\langle\mu\rangle$) is approximately 34. A dedicated sample of 21~$\text{pb}^{-1}$ of low-\pileup data with $\langle\mu\rangle\sim 0.4$ taken in 2017 is used for the \dEdx-to-$\beta\gamma$ calibration. In this data sample, tracks are reconstructed if they have a transverse momentum $\pT > 100~\MeV$ (while the minimum $\pT$ requirement in the standard data sample is $500~\MeV$) allowing the measurement of low-$\beta\gamma$ pions, kaons and protons.
Since LLPs are expected to behave similarly to muons in the calorimeter, data and high-statistic Monte Carlo (MC) samples of $Z\rightarrow\mu\mu$  are used to calibrate the ToF response of the calorimeter.

To optimise the analysis selection, MC samples were produced to simulate events containing long-lived gluinos, charginos,
and staus, with lifetimes ($\tau$) from 3~ns to stable, corresponding to the production diagrams shown in Figure~\ref{fig:diagrams}. These
three signal models are complementary in this study. The gluino samples have large production cross-sections and are suited to probing the high-mass frontier beyond 2~\TeV. The stau samples have production cross-sections that are several orders of magnitude smaller than for gluinos of the same mass, and are suited to probing the few-hundred-\GeV\ mass range.
The chargino sample cross-sections have intermediate values and are useful in probing the mass range around 1~\TeV.

All signal samples were generated using \MGNLO[2.6.2]~\cite{Alwall:2014hca} with up to two additional partons at leading order, and interfaced to \PYTHIA[8.240]~\cite{Sjostrand:2014zea} using the A14 set of tuned parameters (`tune')~\cite{ATL-PHYS-PUB-2014-021}. The \Rhadron samples were generated with the \NNPDF[2.3lo]~\cite{Ball:2012cx} parton distribution function (PDF) set for parton showering and hadronisation, while for the chargino samples the \CTEQ[6.6]~\cite{Pumplin:2002vw} and \MSTW[NLO90CL]~\cite{Martin:2009iq} PDF sets were used, with decays of bottom and charm hadrons performed by \EVTGEN[1.6.0]~\cite{Lange:2001uf}. The CKKW-L merging scheme~\cite{Catani:2001cc,Hoeche:2009rj} was applied to combine the matrix element with the parton shower.

Gluino pair production was simulated for gluino masses ranging from 1.4~\TeV\ to 2.4~\TeV\ within a simplified model inspired by a split-SUSY scenario~\cite{Giudice:2004tc,ArkaniHamed:2004fb}.
The long-lived gluino, which carries colour charge, hadronises to form a colourless composite particle called an \Rhadron.  The details of the \Rhadron simulation are given in Ref.~\cite{ATL-PHYS-PUB-2019-019}. Each gluino decays into a stable neutralino and two quarks via a virtual squark at a very high mass scale in an $R$-parity conserving decay. The gluino acquires a long lifetime as the only decay channel available is via a massive virtual squark. To probe decays with different kinematics, two sets of samples were produced: one with a fixed neutralino mass of $m(\ninoone) = 100~\GeV$, and the other one with a fixed mass splitting of $\Delta m(\gluino,\ninoone) = 30~\GeV$. These two series of mass parameters are complementary and illustrate that the search is open to various models as it does not require explicit decay properties of the charged LLP, like the visible mass of the decay. The nominal cross-section values were calculated at next-to-leading-order (NLO) with resummation of next-to-leading logarithms (NLL). Their uncertainties were taken from an envelope of predictions using different PDF sets and factorisation and renormalisation scales~\cite{Borschensky:2014cia}.

Samples with a combination of chargino--neutralino ($\chinoonepm\ninoone$) and chargino--chargino ($\chinoonep\chinoonem$) events were generated with nearly degenerate chargino and neutralino masses, motivated by the `pure wino' AMSB scenario~\cite{Giudice:1998xp,Randall:1998uk}. Each long-lived chargino decays into a stable neutralino and a pion, where the mass-splitting between the chargino and neutralino is set to 160~\MeV. Although the AMSB model has a specific preference for the chargino's lifetime $(\mathcal{O}(0.2)~\mathrm{ns})$ and mass relation via the loop dynamics~\cite{Ibe:2012sx}, this theoretical constraint was artificially loosened for experimental benchmarking, and charginos with higher lifetimes and masses ranging from $0.7~\TeV$ to $1.4~\TeV$ were examined. A 100\% branching ratio for $\chinoonepm \rightarrow \pi^{\pm}\ninoone$ is assumed. The production cross-sections are computed at NLO plus NLL precision in the limit of mass-degenerate $\chinoonep$, $\chinoonem$, and $\ninoone$,  and with all the other sparticles assumed to be heavy and decoupled~\cite{Fuks:2012qx, Fuks:2013vua}.

Events with pair-produced staus, each of which decays into a $\tau$-lepton and a stable gravitino, were produced in a simplified model motivated by the GMSB scenario~\cite{Alwall:2008ve,Alwall:2008ag,Alves:2011wf}. The stau masses range from 200~\GeV\ to 1~\TeV\ and the small mass of the gravitino is neglected; the long stau lifetime is due to the small coupling to the gravitino.
Signal cross-sections were calculated assuming direct $\stau$ production at NLO in $\alphas$, with soft-gluon emission effects added at NLL accuracy, assuming mass-degenerate left- and right-handed staus ($\stau_\mathrm{L,R}$) with no mixing~\cite{Beenakker:1999xh,Debove:2010kf,Fuks:2012qx,Fuks:2013vua,Fiaschi:2018hgm}.

Inelastic $pp$ interactions were generated using \Pythia[8.186]~\cite{Sjostrand:2007gs} and \EVTGEN[1.6.0] with the \NNPDF[2.3lo] PDF set and the A3 tune~\cite{ATL-PHYS-PUB-2016-017}. The inelastic collisions were overlaid onto the hard-scattering process to simulate the effect of multiple $pp$ interactions. MC samples were reweighed to match the distribution of the mean number of interactions per bunch crossing observed in data.

The MC events were passed through a full detector simulation~\cite{SOFT-2010-01} based on \GEANT~\cite{Agostinelli:2002hh}. The propagation and decays of charginos and staus were simulated within \GEANT, taking into account ionisation loss and interactions with the detector. The propagation of $R$-hadrons and their interactions were handled by \GEANT\ until their decay, at which point the decay chains and subsequent hadronisation were simulated by \PYTHIA[8]; the information about the outgoing particles was then transferred back to \GEANT.


\section{Calibration}
\label{subsec:calibration}

This section describes the calibration of the $\beta\gamma$ measurements provided by the pixel detector and the tile calorimeter.

\subsection{$\beta\gamma$ from ionisation energy loss}
The most probable value (MPV) of the track \dEdxtrunc measured by the pixel detector  varies as a function of the delivered luminosity and detector region.
The radiation dose received, and the consequent charge trapping varies the track \dEdxtrunc by up to 40\% in the data sample. These effects in combination with changing detector operating conditions require a data-derived set of run-by-run and $|\eta|$-dependent corrections such as to equalise the most-probable value of \dEdxtrunc as a function of time and $\eta$ and finally provide the \dEdx value. These corrections are the same as those used in Ref.~\cite{SUSY-2018-42} where their detailed description is given. After these corrections are applied, samples of electrons and muons from selected $Z\rightarrow e e$ and $Z\rightarrow\mu\mu$ events in data show that the corrected \dEdx distribution has negligible dependence on the number of concurrent proton--proton collisions in the event.

The method used to associate a $\beta\gamma$ to a \dEdx value is based on the measurement of low-momentum SM particles.
The correlation between $\beta\gamma$ and \dEdx is extracted from data by fitting a parameterisation of the Bethe-Bloch relation, with the assumptions that $\beta\gamma<$1. Reconstructing tracks with momenta ranging from $100~\MeV$~to a few~$\GeV$ allows to resolve and identify electrons, pions, kaons, protons, and deuterons. The \dEdx spectrum is a superposition of Landau distributions of those particles. Fitting this spectrum extracts the MPV of each particle for each momentum slice. The mapping of $(\beta\gamma, {\mathrm{MPV}}_{\dEdx})$ is redundantly obtained for pion, kaon, proton (and deuteron when statistics are sufficient). While the proton data sample tends to cover the lower $\beta\gamma$ range down to $\beta\gamma\gtrsim0.35$, the pion data sample can cover up to a MIP ($\beta\gamma\approx$3). The kaon data sample overlaps between the two data samples. The mapping used is the one from Ref.~\cite{SUSY-2018-42}, where more details of the $\beta\gamma$ calibration process can be found.

The simulation of the \dEdx response of the pixel detector is based on a realistic charge-deposition model~\cite{WANG20181}, but due to the sensitivity of the \dEdx measurement to detector conditions, including radiation damage, the simulated track \dEdx and especially the probability that a track has a hit in the IBL overflow do not describe the data accurately enough for this analysis. Hence, the \dEdx response for simulated events was modelled by replacing the simulated value with values from a data-driven template~\cite{SUSY-2018-42}, derived from low-momentum events as a function of $\beta\gamma$.

\subsection{$\beta$ from time of flight}

Each TileCal cell provides an independent measurement of $\beta$. The $\beta_i$ measurement of the $i$-th calorimeter cell is obtained from the time measurement in the cell $t_i$ (such that a particle travelling from the interaction point with the speed of light produces a signal at time $t= 0$, in each calorimeter cell), the distance of the cell's centre from the interaction point $l_i$ and the speed of light $c$: $\beta_i = 1 /(1 + \frac{ct_i}{l_i})$. Only cells with an energy deposition above 500~$\MeV$ are considered in order to minimise the effect of the noise contribution.

The final $\betacalo$ exploits the average over $1/\beta_i$ values whose uncertainties are similar to a Gaussian distribution. Thus,  $1/\betacalo$ is obtained as the average over $1/\beta_i$ weighted by $1/\sigma_i^2$, where $\sigma_i$ is the time resolution in the $i$-th cell.

The $\betacalo$ calibration is obtained in consecutive steps: the time offset correction in each cell, the correction based on the track pseudorapidity with respect to the cell's centre, and the determination of the cell time resolution $\sigma_i$. Isolated muons from $Z \rightarrow \mu \mu$ decays are used for this purpose.

The cell time calibration is performed separately for each data-taking year. The core of the time spectrum in each TileCal cell is approximately Gaussian. The mean value of the Gaussian fit in the $\pm$2$\sigma$ region around the peak is taken as the calibration constant to be subtracted from the time measurement. A run-by-run correction did not show any effect on the measured $\betacalo$, hence it is not applied. No effect of the energy deposited in a cell on the mean value of the reconstructed time was visible.%

The reconstructed time in the cells shows a dependence on the distance $\Delta \eta$ of the track's impact point in the cell $\eta_{\mathrm{track}}$ to the cell centre $\eta_{\mathrm{cell}}$ ($\Delta \eta \equiv \eta_{\mathrm{track}}-\eta_{\mathrm{cell}}$). This effect is greater in cells spanning a larger pseudorapidity region. A correction is provided by a linear fit to the mean reconstructed time as a function of $\Delta \eta$ for each cell type\footnote{The cell type is defined by the calorimeter layer and pseudorapidity (A1, BC2, etc), see Figure~\ref{fig:TileCal}.}.

The cell time resolution $\sigma_i$
improves with increasing deposited energy $E$ as
\begin{center}
\begin{equation}\label{math_timeResoVsE}
\sigma_i  = \sqrt{ p_0^2 + \frac{p_1^2}{E} + \left( \frac{p_2}{E} \right) ^2}
\end{equation}
\end{center}

The parameters $p_0, p_1$ and $p_2$ are determined separately in each radial layer (cells A, (B,BC) and D in Figure~\ref{fig:TileCal}) by a fit of Eq.~\ref{math_timeResoVsE} to the muons from $Z\rightarrow\mu\mu$ decays collected in the Run 2 dataset.
For each radial layer, the energy spectrum is divided in 17 slices, and in each slice the time spectrum is fitted with a Gaussian distribution in the $\pm$2$\sigma$ region around the peak. The standard deviation of the fit is taken as the time resolution $\sigma_i$ for the slice with deposited energy $E$. Good agreement between data and the fitting curve is obtained for cells in the two outermost radial layers (BC and D), while worse agreement is obtained for cells in the innermost layer (A cells) due to shorter particle path length and lower energy deposits. For the same reason A cells also show worse time resolution.

\begin{figure}[tbp]
\begin{center}
\subfigure[]{
\label{fig:betacalo_performance_summary}
\includegraphics[align=c,height=65mm]{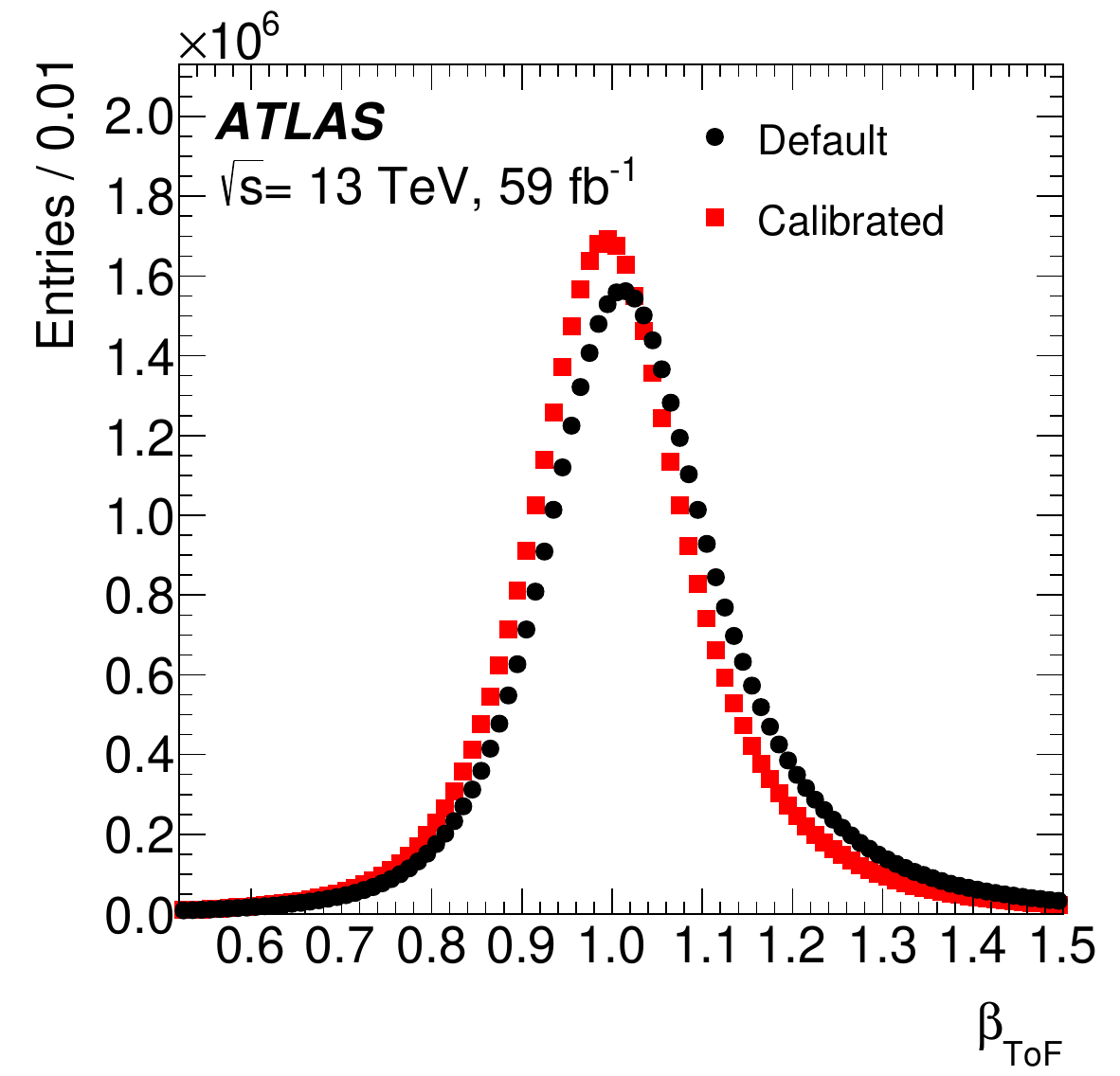}
}
\subfigure[]{
\label{fig:betacalo_resolution_eta}
\includegraphics[align=c,height=65mm]{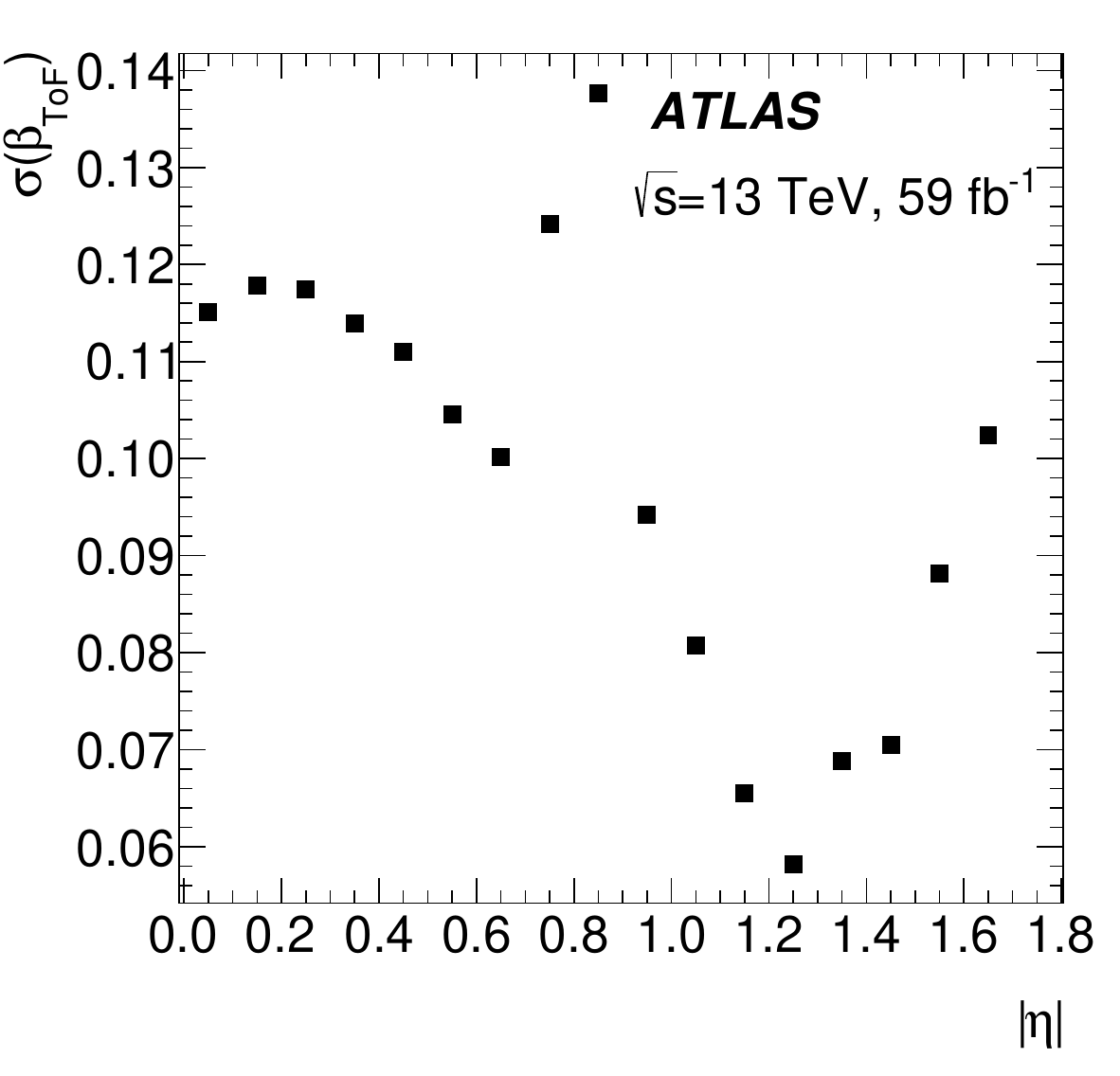}
}
\caption{(a) Distribution of $\betacalo$ obtained with isolated muons
from $Z\rightarrow\mu\mu$ decays (2018 data) with the default calibration (Default) and with the
calibration illustrated in Section~\ref{subsec:calibration} (Calibrated). (b) Dependence of the resolution of $\betacalo$ on the pseudorapidity, using isolated muons from $Z\rightarrow\mu\mu$ decays (2018 data) with all the calibrations applied.}
\label{fig:betacalo}
\end{center}
\end{figure}

The impact of the  calibration described above on the $\betacalo$ performance is checked with isolated muons originating from
$Z\rightarrow\mu\mu$ decays. The performance before and after calibration are compared in Figure~\ref{fig:betacalo_performance_summary} showing a 7\% improvement on $\sigma(\betacalo)$. The $|\eta|$-dependence of the $\betacalo$ resolution is shown in Figure~\ref{fig:betacalo_resolution_eta}. The  $\betacalo$ resolution improves at larger |$\eta$| because of the longer track path. This trend is counterbalanced in the barrel--endcap transition region and in the very high |$\eta$| region as fewer calorimeter cells contribute to the ToF measurement.

\begin{figure}
\centering
\includegraphics[width=0.45\textwidth]{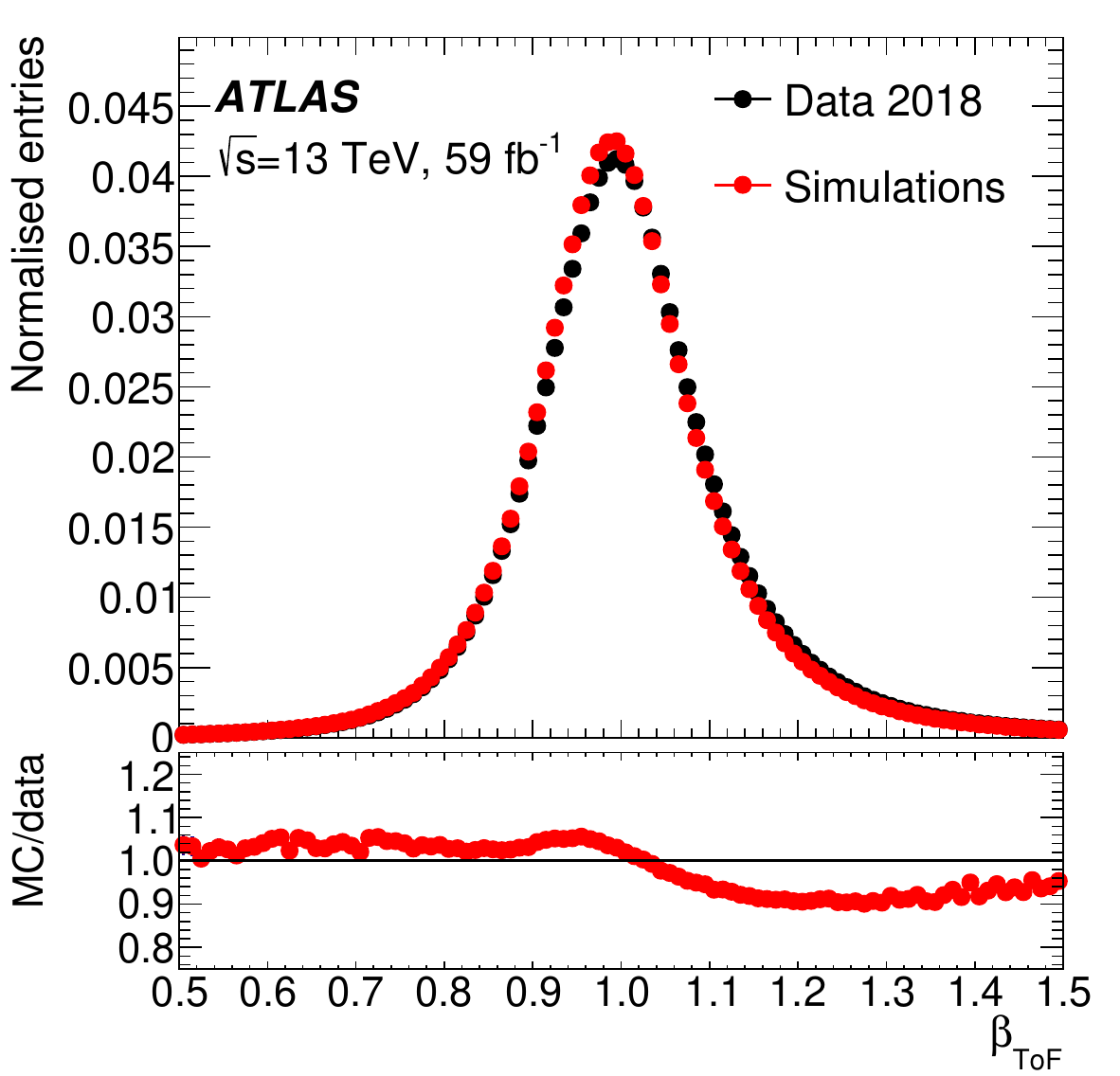}
\caption{Calorimeter $\betacalo$ distribution for $Z \rightarrow \mu \mu$ events in 2018 data and in Monte Carlo after the time smearing procedure. The ratio of the MC to data is shown in the bottom of the plot where a solid black line is drawn at unity for reference.}
\label{fig:betacalo_smearing}
\end{figure}

To account for the difference between the reconstructed time between data and MC, a smearing of the cell time distribution is applied to Monte Carlo samples in such a way as to get the best matching with the $Z \rightarrow \mu \mu$ data taken in 2018.
Figure~\ref{fig:betacalo_smearing} shows the agreement in the calorimeter $\beta$ spectra between $Z \rightarrow \mu \mu$ data and Monte Carlo after the time smearing is applied. The residual difference between data and Monte Carlo is below 5\% for $\betacalo<1$.

\section{Analysis}
\subsection{Overview}
\label{sec:analysis}

This analysis searches for heavy ($m>$200~$\GeV$) charged particles with a proper lifetime $\tau>$3~ns.

Events are selected online using the lowest-threshold un-prescaled calorimetric \MET trigger, which is based on the vectorial energy sum measured in the calorimeters~\cite{TRIG-2019-01}. Further selections are applied to triggered events and candidate tracks as detailed in Section~\ref{subsec:selection}. For the $\beta$-search, a sample of high-momentum isolated tracks with large \dEdx is identified and used to perform two independent measurements, $\massdEdx$ and $\massToF$. For the di-track search, events with two high-momentum, isolated tracks with large \dEdx are selected. For both regions, the search then consists in comparing the data and predicted background yields in trapezoidal mass windows in the [$\massdEdx$, $\massToF$] or in the [$\massAdEdx$, $\massBdEdx$] planes. This enforces that the two measurements in the event -- either two measurements of the same track or of two tracks -- are consistent with the same mass hypothesis. A trapezoidal shape is chosen to take into account the degradation of the mass resolution with increasing mass as well as decreased backgrounds at higher mass.
As described in Section~\ref{subsec:mass_window}, the trapezoidal mass windows are optimised separately for the $\beta$-search and the di-track search; the windows are common to LLPs of the same target mass.

Backgrounds can arise from instrumental effects and tails in the measurements of SM processes, which in the case of the \dEdx include the unavoidable Landau tails of the deposited ionisation energy. The background yield and its distribution in the reconstructed mass spectrum is estimated in a fully data-driven approach, as described in Section~\ref{subsec:background}. Data control samples are used to parameterise the momentum, $\dEdx$ and, when necessary, $\betacalo$ distributions and their interdependence, and then to generate pseudo-data that predict the background distribution in the [$\massAdEdx$, $\massBdEdx$] and the [$\massdEdx$, $\massToF$] planes. Potential signal contamination is minimised in these background samples by inverting some of the selection criteria, and the background estimate method is validated in separate data samples called validation regions.


\subsection{Event selection}
\label{subsec:selection}


%
Events are selected with a trigger based on \MET, which is calculated online using energy measurements in the calorimeter with corrections for multiple $pp$ interactions in each event~\cite{TRIG-2019-01}. The high-level \MET trigger threshold varies from 70~\GeV\ to 120~\GeV\ during the data-taking period depending on the pile-up conditions. The efficiency of signal events to satisfy the trigger ranges from 20\% to over 95\% and depends on the lifetime, mass, and decay mode of the target LLP.

The \MET computation is refined in the offline reconstruction after events are required to satisfy basic data quality selection to ensure all parts of the detector are working correctly~\cite{DAPR-2018-01}. The offline \MET is built from calibrated muons~\cite{MUON-2018-03,MUON-2022-01} and electrons~\cite{EGAM-2018-01,EGAM-2021-02} that satisfy baseline selections, from calibrated jets~\cite{JETM-2018-05} reconstructed at the electromagnetic scale with the \antikt jet clustering algorithm~\cite{Cacciari:2008gp,Fastjet} with radius parameter $R=0.4$, and from a term that includes selected soft tracks not associated with any other objects in the event~\cite{JETM-2020-03} but consistent with the primary vertex (PV).  To remove beam-induced backgrounds and spurious calorimeter signals that could spoil the calculation of \MET, events are rejected if they contain at least one jet tagged as \textit{bad}~\cite{ATLAS-CONF-2015-029} as determined from shower shape information.

In events where the signal LLPs are detector stable (i.e. they decay outside the \mbox{ATLAS} detector), the LLPs leave only modest energy depositions in the calorimeters, even in the $R$-hadron case~\cite{Mackeprang:2010}, and only a fraction of them are reconstructed as a muon owing to their late arrival time in the muon spectrometer. Therefore, most of the momentum of each LLP is not accounted for in the calorimeter or muon system. Any reconstructed jets from initial-state radiation (ISR) or additional partons in the hard scatter provide a visible contribution that results in a measured imbalance of transverse momentum. In events with metastable LLPs (i.e. LLPs that decay inside the ATLAS detector), stable neutralinos or gravitinos will carry away unmeasured momentum that contributes to the measured $\MET$ and increases the trigger efficiency in the assumption that $R$-parity is conserved and the lightest stable sparticle is electrically neutral, as is in the models under consideration, if there is also significant energy deposited in the detector from SM decay products. The $\MET$ trigger and offline selection efficiency increases for metastable LLPs relative to detector-stable LLPs for signals with decay products with significant energy deposition. This includes $R$-hadrons with a light neutralino, but not $R$-hadrons with the fixed and small mass-splitting of $\Delta m(\gluino,\ninoone) = 30~\GeV$, nor the charginos considered here.

For the $\beta$-search, events are required to have $\MET~>~170$~\GeV\ to enhance the signal sensitivity by removing background events from many SM processes. As the di-track search suppresses background with the requirement of two candidate tracks, the offline selection on \MET is relaxed to $\MET~>~20$~\GeV, increasing the acceptance for low-mass LLPs that may produce online \MET due to the presence of an ISR jet in the event but can have low reconstructed offline \MET if both LLP tracks are included in the soft-track term or muon term. The background estimate is fully data-driven and insensitive to the difference between the online and offline \MET thresholds, while the effect on signal efficiency is covered by a dedicated systematic uncertainty, as described in Section~\ref{subsec:systematics}.

After satisfying the trigger and the offline \MET selections, events are required to have a hard-scatter PV with at least two associated reconstructed tracks and to contain at least one ($\beta$-search) or two (di-track search) candidate tracks that satisfy the track-level selections detailed below. In the unlikely case that there are more candidate tracks than required in an event after all selections, the candidate(s) with the highest track $\pt$ is (are) selected. The $\beta$-search and the di-track search are independent and their results are not combined; and an event can then, in principle, be selected by both analyses.

To enrich the selected sample in potential signal events and to minimise background, candidate tracks are required to have $\pt > 120$~\GeV. For masses above 200~\GeV, the LLP production is central and therefore accepting only central-$\eta$ tracks is implemented to maximise significance. For the $\beta$-search, candidate tracks must have $|\eta| <1.6$ to match the $\eta$-acceptance of the central calorimeter, and for the di-track search, candidate tracks must have $|\eta| <1.8$.

Reconstructed tracks must have at least eight clusters across the pixel and SCT detectors. To be considered a candidate, the track must be associated within tolerances with the primary vertex and have an associated cluster in the innermost pixel layer if it passes through an active detector module. Following the optimisation done in Ref.~\cite{SUSY-2018-42} a momentum uncertainty requirement, linearly dependent on $\pt$ and with an upper limit at 200\%, is applied to ensure good mass resolution at low masses and to increase the signal acceptance at high masses, while discarding poorly reconstructed tracks. Similarly, an explicit requirement of at least five associated clusters in the SCT detector is imposed to reject further poorly reconstructed tracks. Additional requirements are applied to candidate tracks to ensure the robustness of the track ionisation computation: no clusters on the track should be consistent with any other track~\cite{PERF-2015-08,PERF-2012-05} and at least two pixel clusters, after discarding the cluster with the highest ionisation, must be included in the \dEdx calculation. Moreover, as the signal is expected to generate isolated tracks, and background processes could acquire significant \dEdx from energy deposits from particles with overlapping trajectories, a track-based isolation requirement is applied. The scalar sum of the $\pT$ of additional primary tracks, in a cone of size $\DeltaR$=0.3 around the candidate track, must be less than 5~\GeV. Although $R$-hadrons are hadrons, the very massive gluino parton produces an expected fragmentation that is very hard~\cite{FAIRBAIRN20071}, and the isolation requirement is above 80\% efficient for the signal models considered.

Additional criteria are applied to reject SM backgrounds from specific processes. To veto tracks from leptonic $W$ decays, the transverse mass of the candidate track must be greater than $130~\GeV$. The transverse mass between the track momentum $\vec{p}_{\mathrm{trk}}$ and $\vec{p}_\mathrm{T}^\mathrm{\,miss}$ is defined as
\begin{eqnarray*}
m_\mathrm{T}(\vec{p}_{\mathrm{trk}},\vec{p}_\mathrm{T}^\mathrm{\,miss}) \equiv \sqrt{2\pt^\mathrm{trk}\,\MET\biggl(1-\cos\Delta\phi(\vec{p}_\mathrm{T}^\mathrm{\,miss}, \vec{p}_{\mathrm{trk}})\biggr)}.
\end{eqnarray*}
For the di-track search, only one of the candidate tracks must satisfy the transverse mass selection.

Tracks from electrons are removed as in Ref.~\cite{SUSY-2018-42} by rejecting a track if any jet with  $\pT > 20~\GeV$ is found within a cone of $\Delta{}R=0.05$ , and has at least 95\% of its energy deposited in the electromagnetic calorimeter. Similarly, SM hadrons are removed by excluding tracks for which any associated jet within a cone of $\DeltaR=0.05$ with  $\pT > 20~\GeV$ has a calibrated energy larger than the track momentum.

Additional selections specific to the $\beta$- and di-track searches are detailed below.
\subsubsection{$\beta$-search selections}
The $\beta$-search requires at least one candidate track that satisfies the selections described in Section~\ref{subsec:selection}. The specific ionisation of the candidate track measured by the pixel detector must be larger than 1.8~$\dEdxunit$, while the most probable value for a MIP is 1.0~$\dEdxunit$ with a resolution of 0.13~$\dEdxunit$. The surviving background, in particular that arising from the  fluctuations in the ionisation tails, is further suppressed by requiring that $\betacalo$ is not compatible with one. This is implemented as $\betacalo < \beta_{\mathrm{cut}}$, where $\beta_{\mathrm{cut}}=1-2\sigma_{\betacalo}$ varies per event and $\sigma_{\betacalo}$ is defined in 0.1-wide slices of $|\eta|$, as shown in Figure~\ref{fig:betacalo_resolution_eta}. The selection has an efficiency for signal events that satisfy the $\dEdx$ requirement ranging from about 80\% at low masses to 95\% at high masses. The efficiency is high because the \dEdx requirement selects signal particles with low $\beta$ that are then likely to satisfy the $\beta_{\mathrm{cut}}$.
The analysis sensitivity is limited to LLPs with lifetimes $\tau \gtrsim 3 \mathrm{ns}$ by the requirement that the track reaches the calorimeter. Conversely, the additional discrimination provided by the calorimeter reduces the background by a factor of about 20 relative to Ref.~\cite{SUSY-2018-42}, which results in an improvement of sensitivity for LLPs with $\tau \gtrsim 10 \mathrm{ns}$.

The efficiency for signal events to satisfy all selections, including the trigger, ranges from 1.0\% to 7.5\% for the simulated LLP events described in Section~\ref{subsec:samples} and with lifetime exceeding 10~$\mathrm{ns}$, and increases with the lifetime of the LLP and its mass. The signal region for the $\beta$-search ($\beta$-SR) is defined in Table~\ref{tbl:CRVR_definition_beta}.

None of the seven tracks associated with the excess reported in Ref.~\cite{SUSY-2018-42} satisfy the $\beta$-search selections.
In particular, none of them have $\betacalo <\beta_{\mathrm{cut}}$. This indicates that the excess identified in Ref.~\cite{SUSY-2018-42} is not due to heavy, highly-ionising and slow particles reaching the hadronic calorimeter.

\subsubsection{Di-track search selections}

The di-track search requires at least two candidate tracks that satisfy the selections described in Section~\ref{subsec:selection}. Additional selections are imposed on properties of the di-track system: the two tracks must have opposite electric charge, and their invariant mass, $m_{\mathrm{inv}}$, calculated with the assumption that the track is a pion, must be larger than 200~\GeV. The latter selection significantly reduces the contribution from $Z$ boson decays, along with other SM sources.

Two signal regions in the di-track search are then defined, differing in the ionisation selections applied to both candidate tracks. In the Discovery Region (Discovery-SR), the potential signal significance is maximised by requiring both tracks to have $\dEdx > 1.7$~$\dEdxunit$, which strongly rejects background. In the Exclusion Region (Exclusion-SR), the exclusion sensitivity is maximised by loosening the ionisation selections to increase the signal acceptance. In this region, one track is required to have $\dEdx > 1.6$~$\dEdxunit$, while the ionisation requirement on the other track is relaxed to $\dEdx > 1.3$~$\dEdxunit$. The efficiency for a signal model with 400~\GeV\ staus with a lifetime of 10~ns to satisfy all selections except the ionisation requirements is about 20\%; in the Discovery-SR and Exclusion-SR, the total efficiency is about 2\% and 5\%, respectively.

\FloatBarrier


\subsection{Background estimation}
\label{subsec:background}


%

%

The mass distribution of tracks from background processes in the signal regions is estimated by using a data-driven technique in which tracks are sampled from several control regions (CR). The mass of a track is defined by its momentum and $\beta\gamma$; therefore, the background mass distributions, $\massdEdx$  and $\massToF$,  are constructed by sampling templates of these variables extracted from control regions. Validation regions (VR) are defined to verify the closure of the analysis method.

For a narrow $\Delta\eta$ slice, \dEdx and $\pT$, as well as $\betacalo$ and $\pT$, are assumed to be uncorrelated for background tracks. In addition to the VRs, cross-checks on both assumptions are performed and described in Section~\ref{subsec:systematics}. The independence of \dEdx and $\pT$ for minimum ionising particles is explicitly confirmed using muon tracks from an independent sample of selected $Z\rightarrow\mu\mu$ events.

The $\pT$ distribution of background tracks, for each $|\eta|$ slice, is sampled from kinematic CRs and used as a template for the background track momentum in the signal and validation regions. The kinematic CRs are defined by inverting the \dEdx selection for candidate signal tracks. The \dEdx and the $\betacalo$ distributions are sampled from additional CRs, as described in the sections below.

To generate a `pseudo-data' background track, a pair of $\pt$ and $|\eta|$ values is sampled from the kinematic CR template. A \dEdx (or a $\betacalo$ ) value is sampled from the corresponding $|\eta|$ bin of the \dEdx (or $\betacalo$) template. From these sampled values, the track mass, $\massdEdx$ or $\massToF$, is calculated using the $\dEdx\mbox{--}\beta\gamma$ calibration (or the $\betacalo$). Enough tracks are generated that the number of pseudo-data samples does not limit the accuracy of the predictions. The normalisation and validation of each background estimate is described in the following sections.

\subsubsection{$\beta$-search background estimation and validation}
Two control regions, kin-CR and $\beta\gamma$-CR, are defined adjacent in phase space to the signal region (SR) (see Table~\ref{tbl:CRVR_definition_beta}). The kin-CR is defined by inverting the \dEdx requirement and relaxing the $\betacalo$ requirement used in the SR, and the $\beta\gamma$-CR is defined by inverting the $\met$ requirement used in the SR and removing the $\dEdx$ and relaxing the $\betacalo$ requirements. Data events in the $\beta\gamma$-CR are reweighed with an $\MET$ trigger threshold weight to prevent the $\MET$ trigger differences causing a bias on the background estimate. The \dEdx and the $\betacalo$ distributions  in the $\beta\gamma$-CR serve as the  template distributions for the background mass prediction of $\massdEdx$ and $\massToF$, respectively. Only tracks with $\betacalo<1$ are considered as this is the condition to calculate $\bgToF$ and then $\massToF$.

\begin{table}[tp]
\centering
\caption{Definitions of the signal, control and validation regions for the $\beta$-search. The $\beta_{\mathrm{cut}}$ varies per event and is defined as $\beta_{\mathrm{cut}}=1-2\sigma_{\betacalo}$. The estimation of the background for each signal and validation region requires two dedicated control regions.}
\scriptsize
\begin{tabular}{|c|c|c|c|}
\hline{$\beta$-search region} & $\met~\mathrm{[\GeV]}$ &  $\dEdx~\mathrm{[\dEdxunit]}$  &  $\betacalo$ \\
\hline
$\beta$-SR                                   & $> 170$ &  $> 1.8 $     & $< \beta_{\mathrm{cut}}$ \\
kin-CR                             &              $> 170$  &   $< 1.6 $    & $< 1.0$ \\
$\beta\gamma$-CR         &         $<150$                    &   --               & $< 1.0$ \\

\hline
High-$\betacalo$-VR                                     & $> 170$  &  $> 1.8 $      & [$\beta_{\mathrm{cut}}$, 1.0] \\
High-$\betacalo$-VR kin-CR                      &  $> 170$ &  $< 1.6 $       & [$\beta_{\mathrm{cut}}$, 1.0] \\
High-$\betacalo$-VR $\beta\gamma$-CR   &              $<150$               &           --         & [$\beta_{\mathrm{cut}}$, 1.0] \\
\hline

Low-\dEdx-VR                            & $> 170$  &  [1.05,1.6]     & $< \beta_{\mathrm{cut}}$ \\
Low-\dEdx-VR kin-CR                      & $> 170$  &  $< 1.05 $     & $< 1.0$ \\
Low-\dEdx-VR $\beta\gamma$-CR   &         $<150$                    &   $< 1.6 $      & $< 1.0$ \\
\hline

\end{tabular}
\label{tbl:CRVR_definition_beta}
\end{table}%

Pseudo-data background tracks are simulated using a pair of $1/\pt$ and $|\eta|$ values sampled from the kinematic control region template and a \dEdx (or a $\betacalo$ ) value sampled from the corresponding $|\eta|$ bin of the \dEdx (or $\betacalo$) template. From these sampled values, the track mass $\massdEdx$ and $\massToF$ are calculated. Finally, the pseudo-data samples are normalised to data in a sub-region of kin-CR that is expected to be fully depleted in signal, with $\betacalo > \beta_{\mathrm{cut}}$ and both $\massdEdx$  and $\massToF$ lower than $160~\GeV$.

The procedure for estimating both the normalisation and shape of the expected background is validated  in two regions.  One validation region is characterised by high-$\betacalo$ (High-$\betacalo$-VR), i.e. it contains tracks that do not satisfy the $\betacalo$ cut used for the signal selection (i.e. $\betacalo>\beta_{\mathrm{cut}}$) but extends to high \dEdx.
The other validation region is characterised by low-$\dEdx$ (Low-\dEdx-VR), i.e. it contains tracks that do not satisfy the \dEdx requirement used for the signal selection (track ionisation in the range of $[1.05,1.6]~\dEdxunit$) but tests low values of \betacalo. The definition of these regions and the corresponding control regions used for the background estimation are shown in Table~\ref{tbl:CRVR_definition_beta}.
The validation regions are mutually exclusive and exclusive with the signal region by construction.
The availability of the $\betacalo$ measurement allows to define validation regions without a \pt upper limit and therefore allows the background estimate to be validated at high masses, which is an improvement with respect to the validation strategy used in the previous \dEdx analysis~\cite{SUSY-2018-42}.

The contribution of possible signal contamination in each validation region was studied by comparing the number of signal events from various signal samples to the number of background tracks predicted by the background estimation procedure. It was found that the possible signal contamination was  smaller than about~5$\%$ ($\text{N$_{\text{signal}}$}/\sqrt{\text{N$_{\text{background}}$}}$<0.9) in the Low-\dEdx-VR and smaller than about~12$\%$ ($\text{N$_{\text{signal}}$}/\sqrt{\text{N$_{\text{background}}$}}<1.2$)  in the High-$\betacalo$-VR for all samples and masses not excluded by previous searches. The signal contamination is found to be negligible within all control regions and regions used for normalisation.

\begin{figure}
\begin{center}
\subfigure[]{
\label{fig:masscalo_lowdEdxVR}
\includegraphics[width=0.45\textwidth]{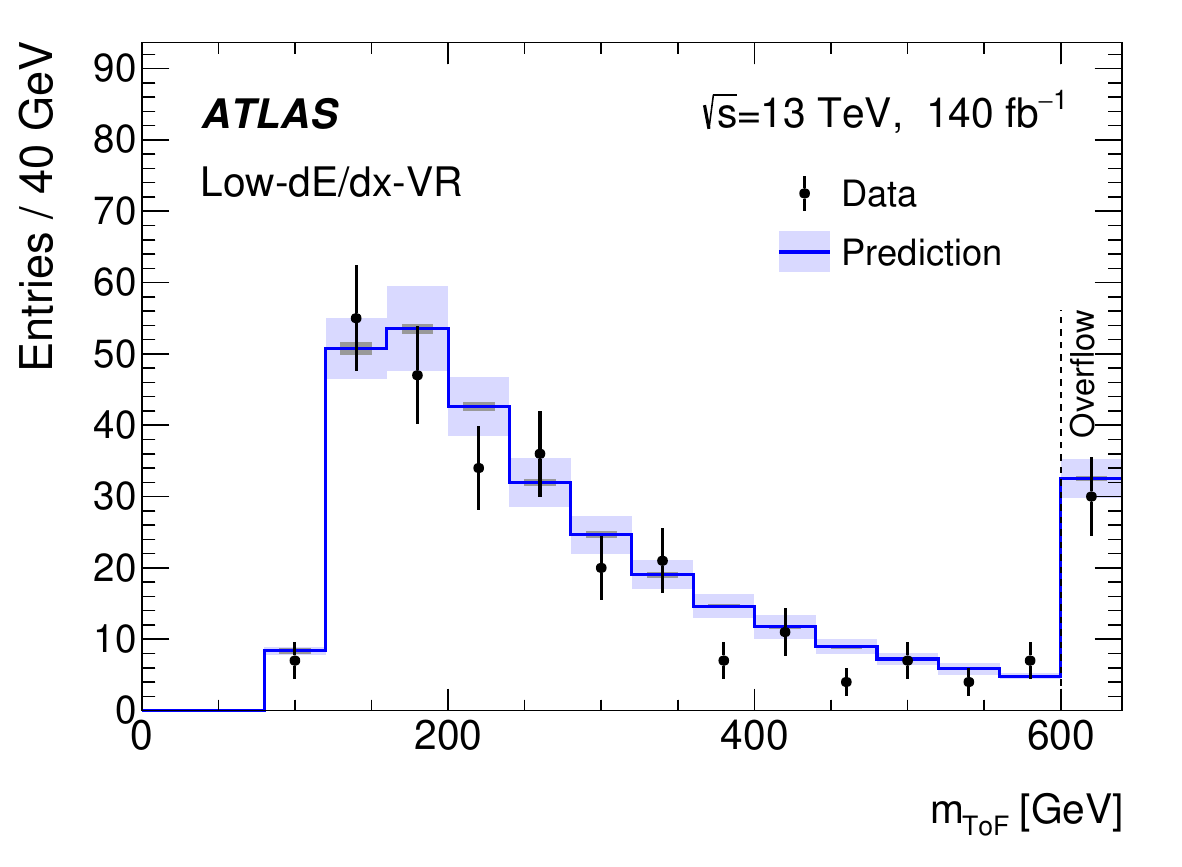}
}
\subfigure[]{
\label{fig:massdEdx_lowdEdxVR}
\includegraphics[width=0.45\textwidth]{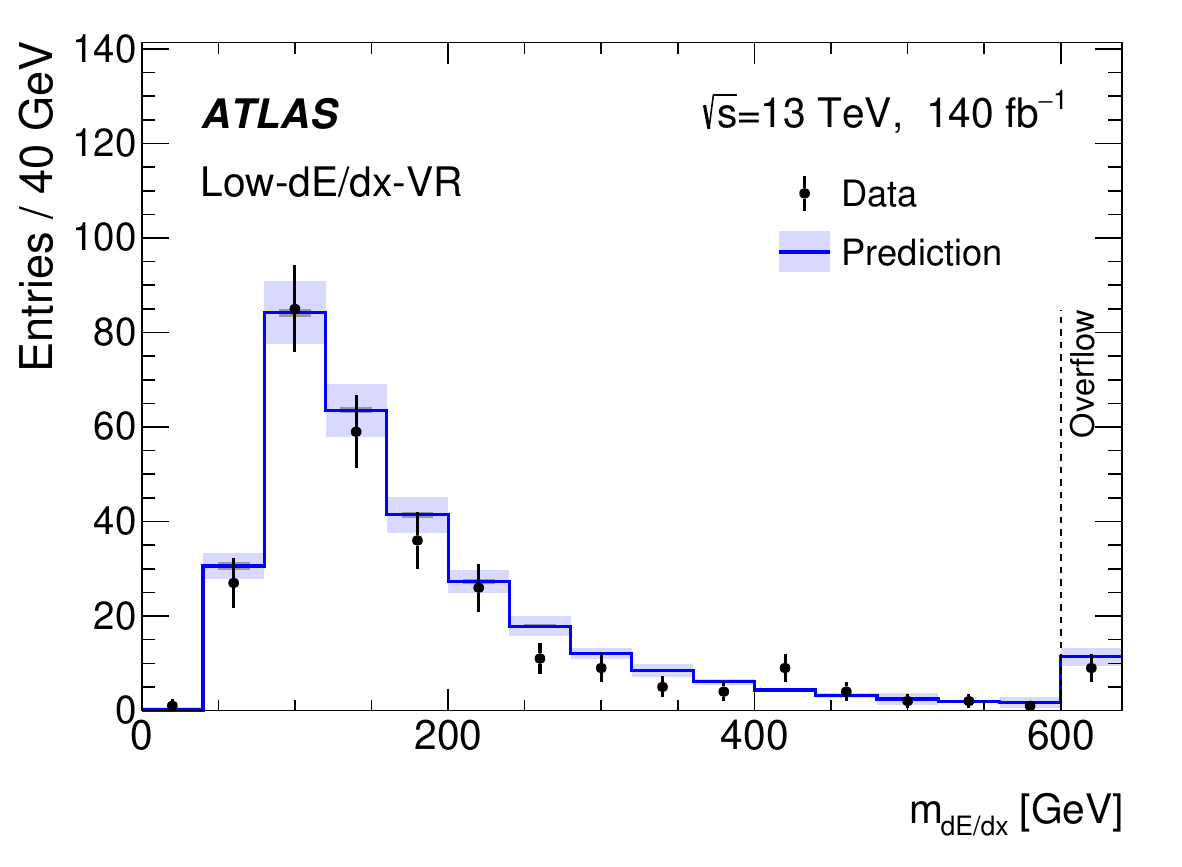}
}
\caption{Comparison of predicted (a) $\massToF$ and (b) $\massdEdx$ background to data in the low-\dEdx validation region. The statistical and systematic uncertainty in the predicted background is calculated as indicated in Section~\ref{subsec:systematics} and shown as a coloured band. The histogram overflow is added into the rightmost bin.}
\label{fig:mass_cfr_lowdEdxVR}
\end{center}
\end{figure}

\begin{figure}
\begin{center}
\subfigure[]{
\label{fig:masscalo_highBetaCaloVR}
\includegraphics[width=0.45\textwidth]{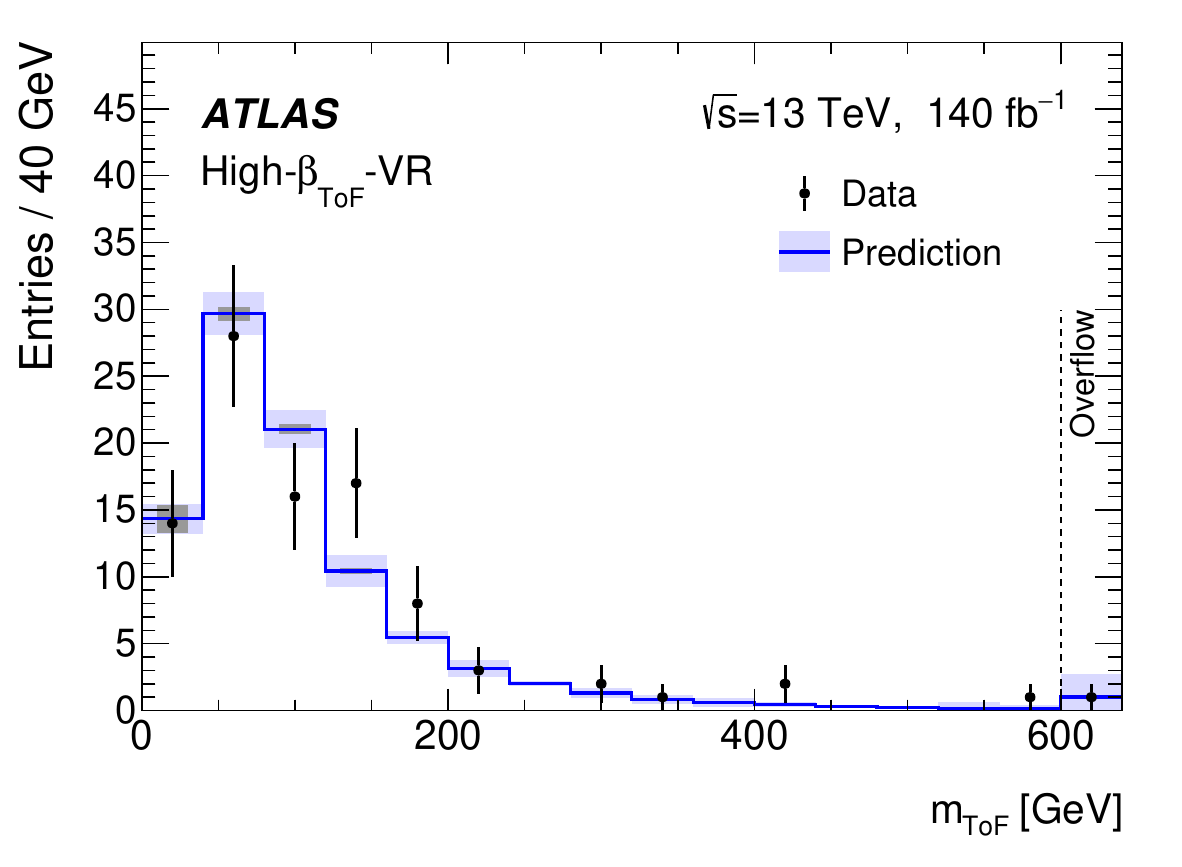}
}
\subfigure[]{
\label{fig:massdEdx_highBetaCaloVR}
\includegraphics[width=0.45\textwidth]{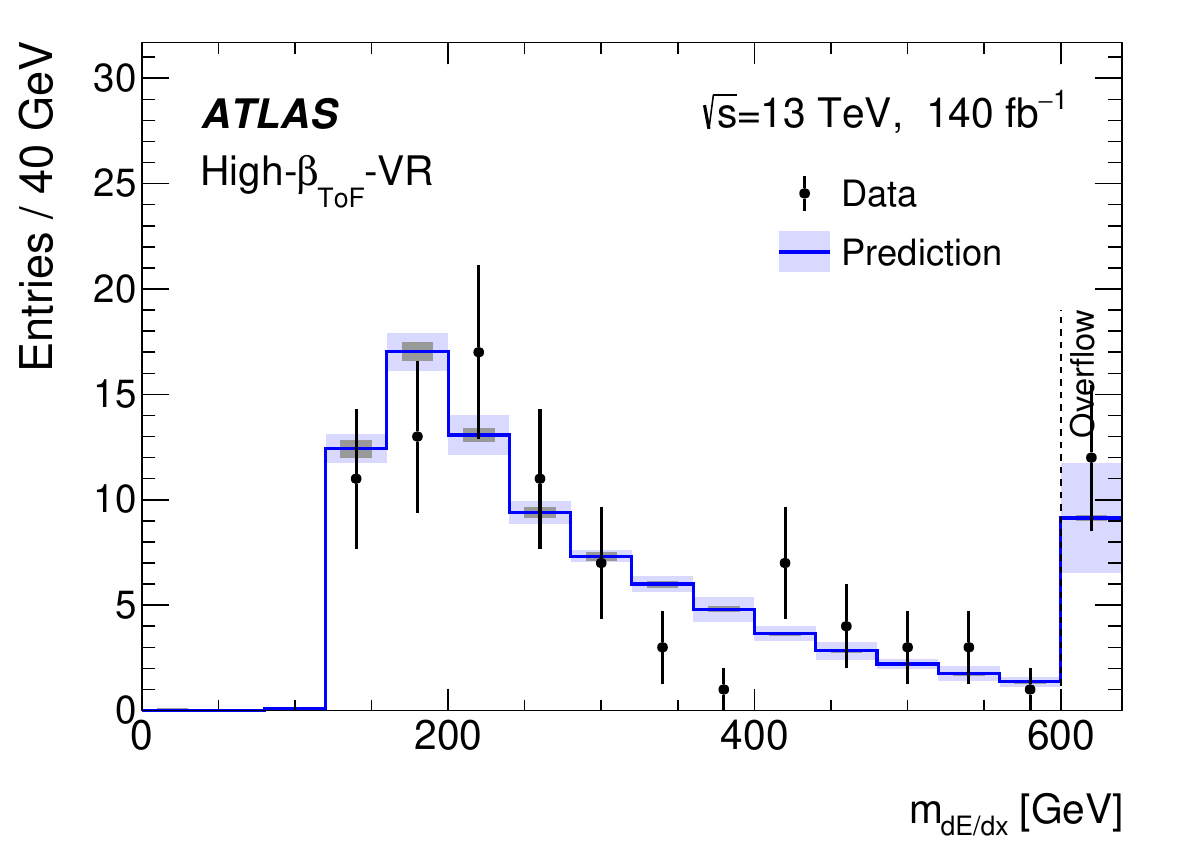}
}
\caption{Comparison of  predicted (a) $\massToF$ and (b) $\massdEdx$ background to data in the high-$\betacalo$ validation region. The statistical and systematic uncertainty in the predicted background is calculated as indicated in Section~\ref{subsec:systematics} and shown as a coloured band. The histogram overflow is added into the rightmost bin.}
\label{fig:mass_cfr_highBetaCaloVR}
\end{center}
\end{figure}

The predicted $\massToF$ ($\massdEdx$) background distribution compared with the $\massToF$ ($\massdEdx$) distribution in the data are shown in Figure~\ref{fig:mass_cfr_lowdEdxVR} and \ref{fig:mass_cfr_highBetaCaloVR}.
There is good agreement between prediction and data both in the shape of the distributions and in the total yield. The predicted yield, including statistical and systematic uncertainties, is 316$\pm27$ (91$\pm6$) while the observed yield is 290 (93) for the Low-$\dEdx$-VR (High-$\betacalo$-VR).

\FloatBarrier

\subsubsection{Di-track background estimation and validation}

The number of events from background processes with two tracks that satisfy all selections and the $\massdEdx$ distribution of each track is predicted from pseudo-data events sampled from templates constructed from control regions. The kinematics of pairs of background tracks are extracted from a sample of events (kin-CR) that have two tracks that satisfy an inverted $\dEdx$ selection. Sampling two tracks from the same event ensures that all kinematic correlations are retained. Inverting the $\dEdx$ selection on both tracks ensures that there is no significant signal presence in the control region.

The $\dEdx$ template is formed from events that have two tracks, at least one of which has a low $\pt$ value to exclude possible signal ($\dEdx$-CR). As there is a small residual correlation between $\dEdx$ and $|\eta|$, the $\dEdx$ template is binned in $|\eta|$. The definitions of the signal and control regions, as well as the validation regions defined below, are shown in Table~\ref{tbl:CRVR_definition_ditrack}.

\begin{table}[tp]
\centering
\caption{Definitions of the signal, control and validation regions for the di-track search. Events must satisfy all event-level requirements and have at least two candidate tracks that satisfy all other selection requirements, as defined in the text. Tracks are ranked first by $\pt$ and then by $\dEdx$; Track 1 is the candidate track with the larger $\pt$ or $\dEdx$, depending on the stage of the selection. The estimation of the background for each signal and validation region requires two dedicated control regions.}
\scriptsize
\begin{tabular}{|c|c|c|c|c|c|c|}

\hline
Di-track region & \makecell{$m_{\mathrm{inv}}$\\$~\mathrm{[\GeV]}$} & \makecell{$m_{\mathrm{T}}$\\$~\mathrm{[\GeV]}$} & \makecell{Track 1 $\pt$\\$~\mathrm{[\GeV]}$}&   \makecell{Track 2 $\pt$\\$~\mathrm{[\GeV]}$} & \makecell{Track 1 $\dEdx$\\$~\mathrm{[\dEdxunit]}$}  &  \makecell{Track 2 $\dEdx$\\$~\mathrm{[\dEdxunit]}$} \\
\hline
Exclusion-SR    & \multirow{4}{*}{$> 200$}  &  \multirow{4}{*}{$> 130 $} &           $> 120$            &              $> 120$                 & $>1.6$ & $>1.3$ \\
Discovery-SR    &                                         &                                           &            $> 120$             &            $> 120$                   &  $>1.7$ & $>1.7$ \\
kin-CR             &                                         &                                           &            $> 120$              &            $> 120$                  & $<1.3$ & $<1.3$ \\
$\dEdx$-CR           &                                         &                                          &           $>50$                   & $[50,120]$                           & --          &  --          \\
\hline
LowpT-VR         &  \multirow{3}{*}{$> 200$} &\multirow{3}{*}{$> 130$}     &   $[70,100]$                   & $[70,100]$                            & $>1.6$ & $>1.3$  \\
LowpT-VR kin-CR &                                   &                                             &   $[70,100]$                   & $[70,100]$                            & $<1.3$ & $<1.3$  \\
LowpT-VR $\dEdx$-CR &                                &                                             &         $>10$                    & $[10,70]$                               & --          & --             \\
\hline
W-VR              &    \multirow{3}{*}{$< 200$}  &\multirow{3}{*}{-}              &   $[70,100]$                   & $[70,100]$                            & $>1.6$ & $>1.3$  \\
W-VR kin-CR &                                           &                                           &   $[70,100]$                   & $[70,100]$                            & $<1.3$ & $<1.3$  \\
W-VR dEdx-CR &                                         &                                           &         $>10$                    & $[10,70]$                              & --          & --             \\
\hline
InvMass-VR              &    \multirow{3}{*}{$< 200$}  &\multirow{3}{*}{$> 130$}  &  $> 120$                 & $> 120$                    & $>1.6$ & $>1.3$  \\
InvMass-VR kin-CR &                                           &                                      &   $> 120$                   & $> 120$                         & $<1.3$ & $<1.3$  \\
InvMass-VR $\dEdx$-CR &                                         &                                      &          $>50$              & $[50,120]$                     & -          & -             \\
\hline
Z-VR              &    \multirow{3}{*}{$[80,100]$}  &\multirow{3}{*}{-} &  $> 120$                   & $> 120$                    & $>1.6$ & $>1.3$  \\
Z-VR kin-CR &                                           &                                           &   $> 120$                   & $> 120$                         & $<1.3$ & $<1.3$  \\
Z-VR $\dEdx$-CR &                                         &                                           &          $>50$              & $[50,120]$                     & --          & --             \\
\hline
\end{tabular}
\label{tbl:CRVR_definition_ditrack}
\end{table}%

To form a pseudo-data event, an event from the kin-CR is assigned two independent $\dEdx$ values sampled from the corresponding $|\eta|$ bins of the $\dEdx$ template, and the track masses $\massAdEdx$ and $\massBdEdx$ are calculated. Events from the kin-CR can be reused with different sampled $\dEdx$ values.

Enough pseudo-data events are generated so that the statistical uncertainty due to sampling is negligible. Treatment of the statistical correlation due to the kin-CR event reuse is discussed in Section~\ref{subsec:systematics}. Each pseudo-data event is normalised to data by the relation: $\frac{N_{\mathrm{kin}}}{N_{\mathrm{pseudo}}}\times\frac{1}{(1-f_1)(1-f_2)}$, where $N_{\mathrm{kin}}$ is the number of events in the kinematic template, $N_{\mathrm{pseudo}}$ is the number of pseudo-data events, and $f_1$ and $f_2$ are the fraction of tracks in the $\dEdx$ template with high $\dEdx$ in the $\eta$-bin corresponding to the first and second track of the pseudo-data event, respectively. The $(1-f_1)$ and $(1-f_2)$ factors correct for the exclusion of high $\dEdx$ events from the kin-CR.  Applying the signal $\dEdx$ selection on the pseudo-data events produces a normalised distribution of $\massdEdx$ for two-track events expected in the signal region.

\begin{figure}
\begin{center}
\subfigure[]{
\label{fig:mass_lowpt}
\includegraphics[width=0.45\textwidth]{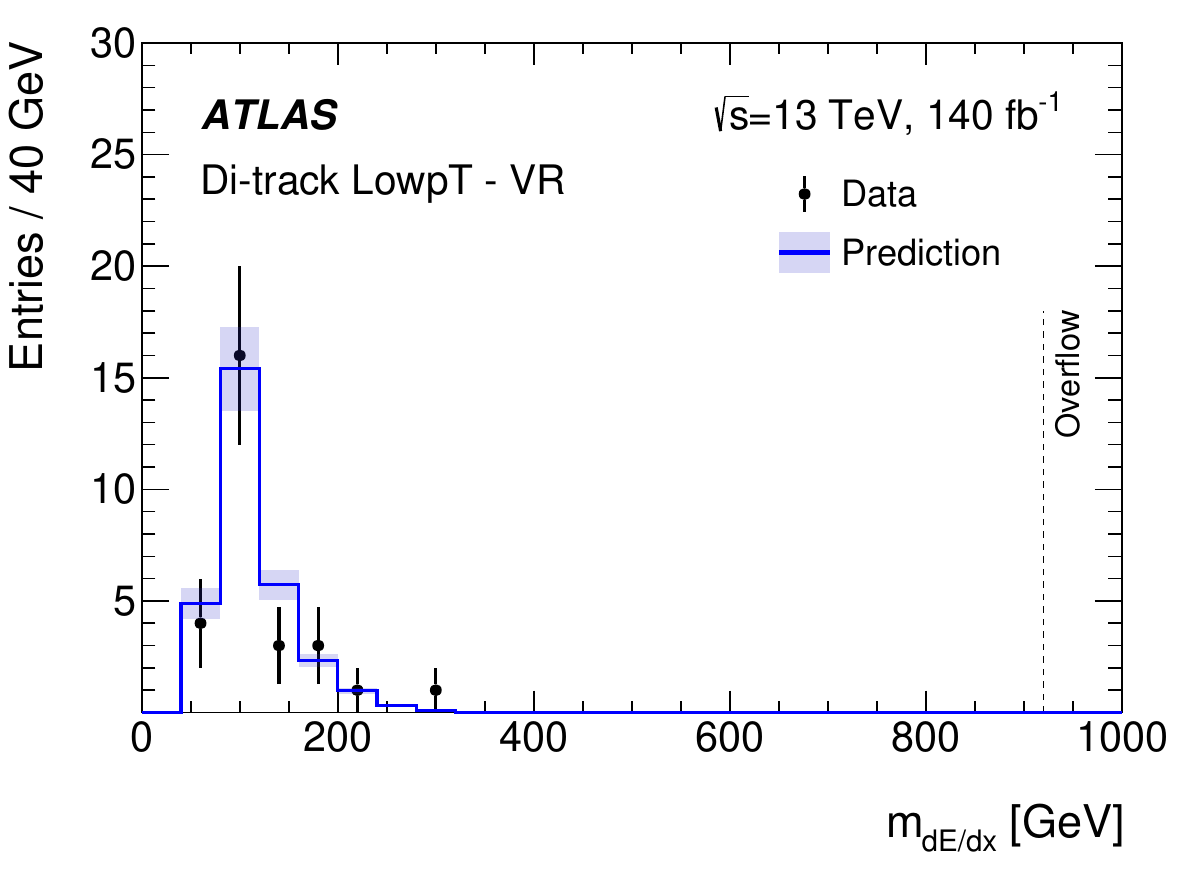} }
\subfigure[]{
\label{fig:mass_inv_inv}
\includegraphics[width=0.45\textwidth]{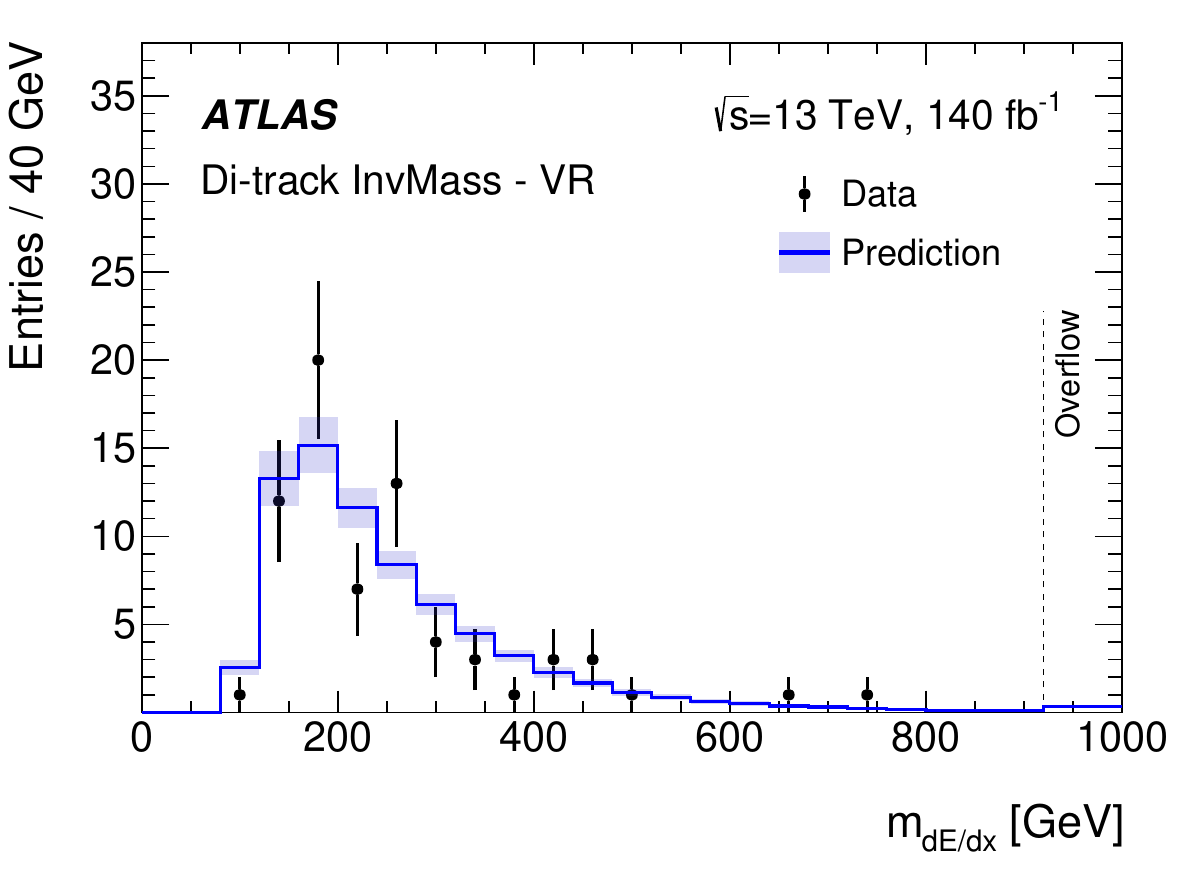} }
\subfigure[]{
\label{fig:mass_WVR}
\includegraphics[width=0.45\textwidth]{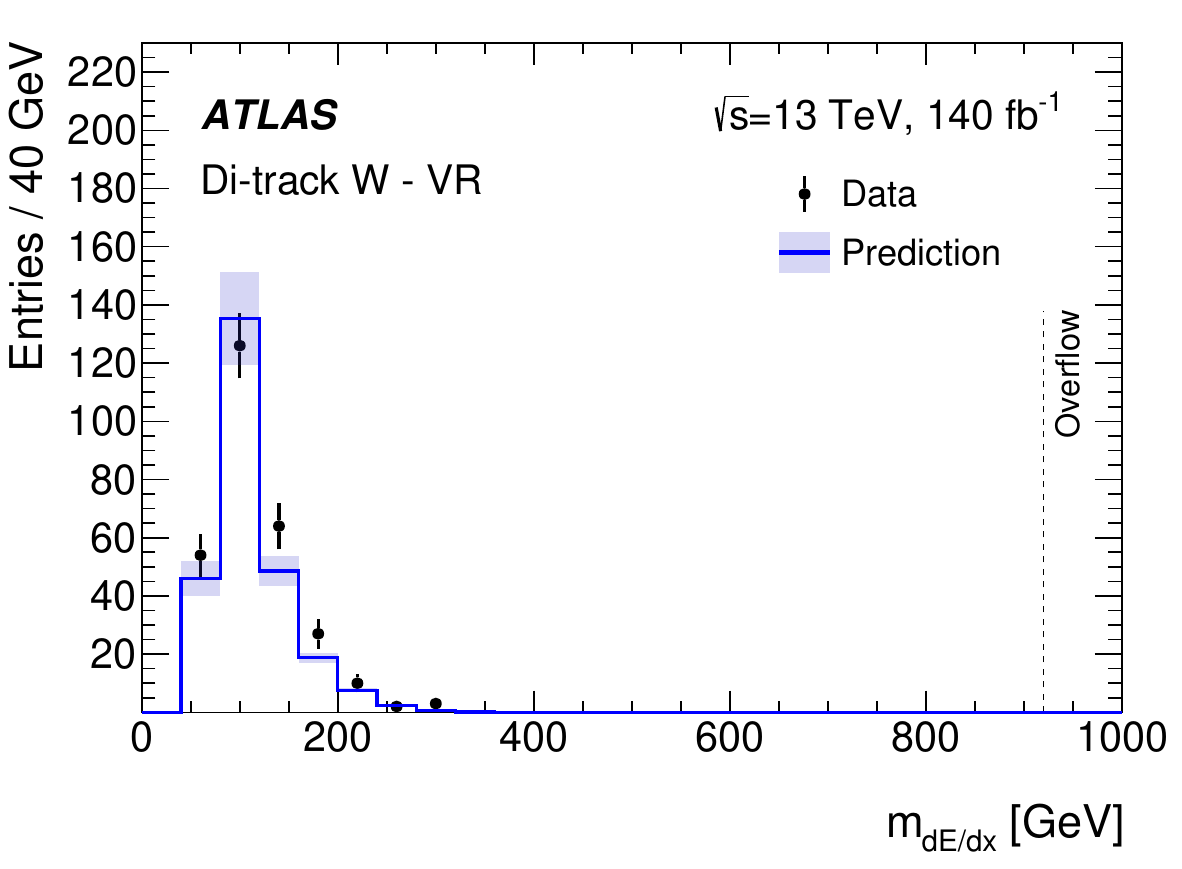} }
\subfigure[]{
\label{fig:mass_Zmumu}
\includegraphics[width=0.45\textwidth]{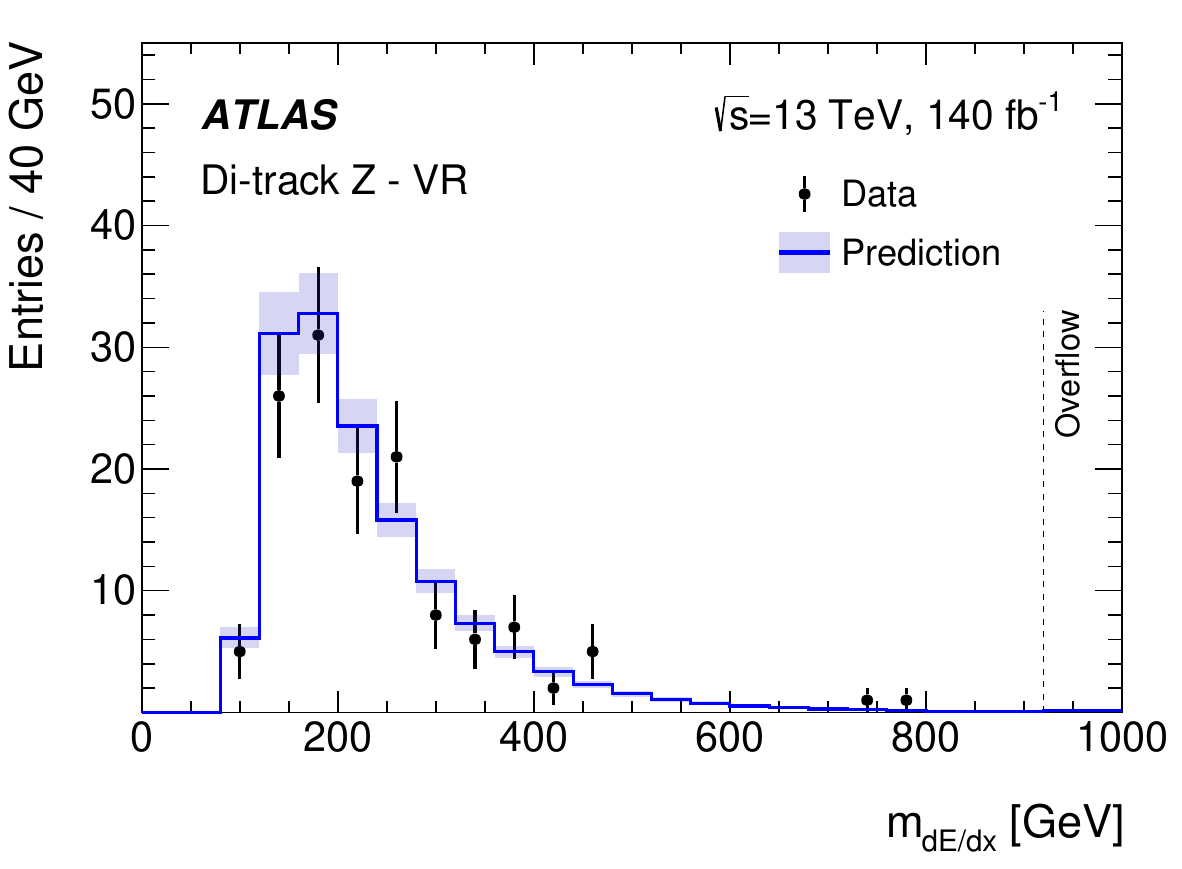} }
\caption{Comparison of the predicted $\massdEdx$ background to data in the di-track validation regions. The statistical and systematic uncertainty in the predicted background is calculated as indicated in Section~\ref{subsec:systematics}.}
\label{fig:mass_ditrack_VR}
\end{center}
\end{figure}

The background estimation procedure is validated in several regions, in which the full pseudo-data method is repeated using dedicated control regions for each validation region, as defined in Table~\ref{tbl:CRVR_definition_ditrack}. The inclusive LowpT-VR keeps all selections the same as the signal region, except for the track $\pt$ requirements, which are loosened so that each signal track is required to have $\pt$ between 70 and 100~\GeV. The $\dEdx$ control region for LowpT-VR similarly loosens the lower track $\pt$ to 10~\GeV, in order to ensure sufficient template statistics. To ensure that the method closes at nominal values of track $\pt$, an independent InvMass-VR is constructed, in which the track $\pt$ requirements are the same as the signal region, but the invariant mass requirement of the two tracks is inverted. Additionally, two higher statistics validation regions are formed that target background processes with $W$ and $Z$ bosons; both regions omit the one-track transverse mass selection, which dramatically enhances the number of $W$ bosons relative to the InvMass-VR for the W-VR, and the Z-VR additionally omits the offline $\met$ selection and requires that the two tracks each be matched to a reconstructed muon and have an invariant mass consistent with the $Z$ boson. Good agreement between the pseudo-data prediction and the observed data is seen in all validation regions, as shown in Table~\ref{tab:VR_yields_ditrack} and Figure~\ref{fig:mass_ditrack_VR}.

\begin{table}[tp]
\centering
\caption{The expected and observed yields in the di-track validation regions. Only the statistical uncertainty is included in the prediction. Not all regions are orthogonal}
\scriptsize
\begin{tabular}{|c|c|c|}
\hline
Region & {Predicted yield} & {Observed yield} \\
\hline
LowpT-VR  & 13.0$\pm2.6$ & 14 \\
InvMass-VR  & 36$\pm1.9$ & 35 \\
W-VR & 126$\pm4.0$ & 138 \\
Z-VR & 71$\pm4.7$ & 66 \\
\hline
\end{tabular}
\label{tab:VR_yields_ditrack}
\end{table}

An additional validation of the expected yield in the signal and validation regions was performed using an $ABCD$ method~\cite{EXOT-2012-10} that requires the leading track to satisfy all signal selections and uses the $\pt$ and $\dEdx$ of the second track as the independent variables. Yields were found to be in agreement with both the pseudo-data model prediction and the observed data in the validation regions. To validate the background behaviour at higher $\dEdx$ values needed for the Discovery-SR, the $ABCD$ estimate was tested in a modified InvMass-VR and Z-VR with symmetric $\dEdx$ selections of 1.3, 1.4, 1.5, and 1.6$~\mathrm{\dEdxunit}$, and agreement was found within statistical uncertainty in every test. To obtain enough statistics to test the behaviour at $\dEdx > 1.7~\mathrm{\dEdxunit}$, a low-$\pt$ single-track validation region was developed, with a prediction of $1204\pm38$ events with $1246$ observed in data.

\FloatBarrier


\subsection{Mass window definition}
\label{subsec:mass_window}

The sensitivity to signal LLPs is measured comparing data, expected
background, and potential signal in two-dimensional mass windows.

\begin{figure}[tp]
\resizebox{0.70\textwidth}{!}{
\includegraphics{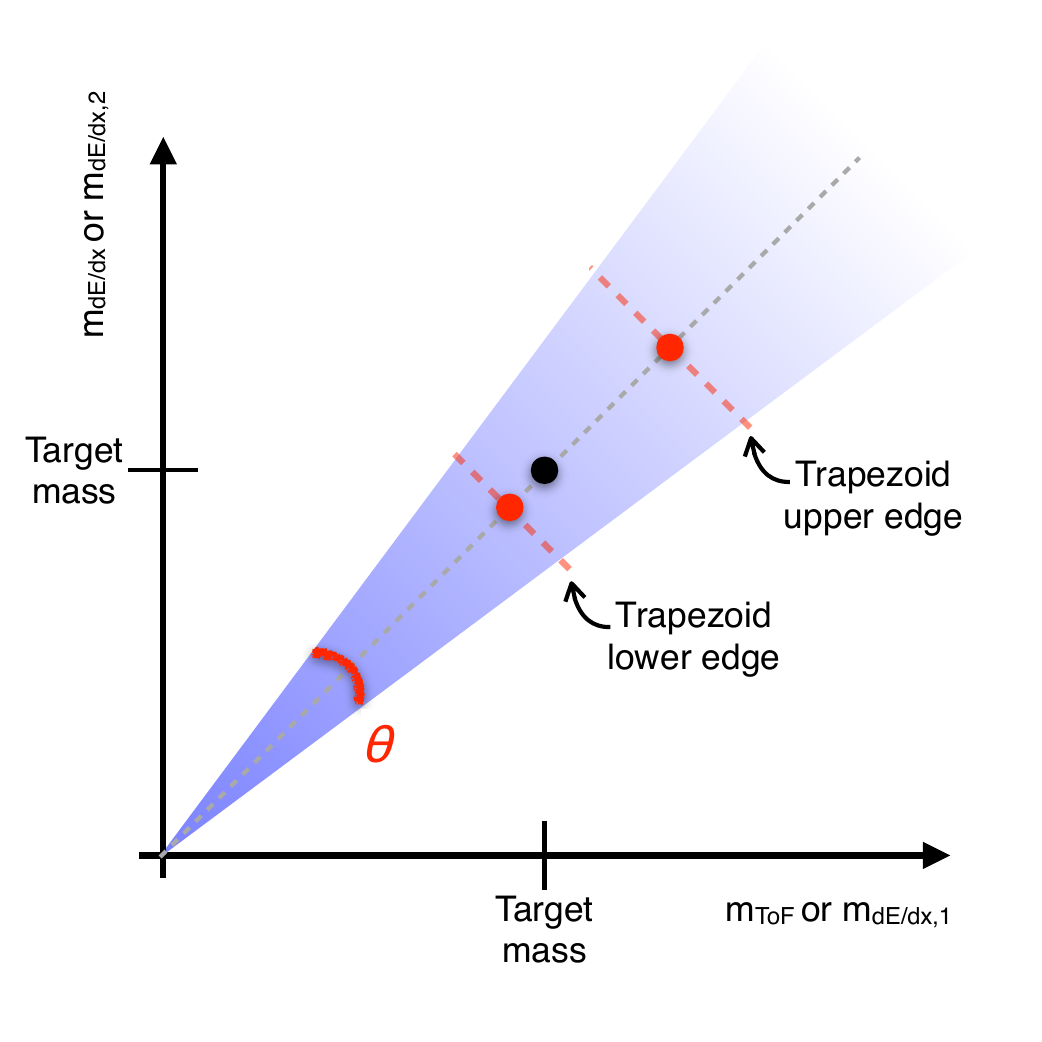}
}
\centering
\caption{The trapezoidal mass windows are determined within an angle
$\theta$ in the [$\massToF$, $\massdEdx$] or [$\massAdEdx$,$\massBdEdx$] plane. The mass points at the centre of the upper and lower edges of the trapezoids are shown as red dots and parameterise the mass window for each target mass. The target mass of the trapezoid is shown as a black dot.}
\label{fig:trapezoid_definition}
\end{figure}

Two masses, $\massdEdx$ and $\massToF$, can be calculated for each track
surviving the $\beta$-driven signal selection using $\bgdEdx$, $\bgToF$
and the momentum, and similarly two $\massdEdx$ are calculated in
each signal di-track event, one per track. The measurement of two masses allows to define
surfaces in the [$\massdEdx$, $\massToF$] or
[$\massAdEdx$, $\massBdEdx$] plane that are optimised for each signal
mass. A trapezoidal shaped window is found to
maximise sensitivity, reflecting the worsening mass resolution at
higher mass, caused primarily by the worsening momentum resolution at
higher momenta. The opening angle limiting the trapezoids is assumed to be symmetric around the line
$\massdEdx$ = $\massToF$ or $\massAdEdx$ = $\massBdEdx$ and is
sketched in Figure~\ref{fig:trapezoid_definition}.

\subsubsection{$\beta$-search mass windows}
The mass windows are constructed to contain about~70\% of the expected
signal. The selected surfaces are different for different signal masses. The trapezoidal mass windows are determined within an angle in the [$\massdEdx$, $\massToF$] plane that is determined by the spread in $\beta\gamma_{\dEdx}$ relative to $\beta\gamma_{\mathrm{ToF}}$. The spread in the difference between the two $\beta\gamma$ measurements translates into an expected spread in $\massdEdx$ relative to $\massToF$. As $\massdEdx$ and $\massToF$ are both determined using the same momentum measurement, their difference for each track reflects only the difference in $\beta\gamma$.
The trapezoid opening angle is set to 22 degrees around the bisector of the [$\massdEdx$, $\massToF$] plane, approximately optimal for all mass points studied.
The trapezoids do not depend nor on the LLP lifetime nor on their identity, but only on their mass. The trapezoid lower and upper edges (see Table~\ref{tab:h_trapezoid_mass_windows}) are defined through a procedure designed to optimise the signal sensitivity. First, the lower edge is scanned to find the optimal sensitivity with the upper edge fixed at 7~\TeV. Then, the upper edge is defined by lowering it until the sensitivity is maximal while still keeping at least 70\% of the signal events.

\begin{table}[tp]
\centering
\caption{The 2D mass window parameters for the $\beta$-driven
signal region derived using the trapezoidal method, given an opening angle of 22 degrees. The lower edge mass identifies the mass point ($\massdEdx$, $\massToF$) at the centre of the trapezoidal lower edge. The upper edge mass locates the centre at point ($\massdEdx$, $\massToF$) of the trapezoid upper edge in the 2D mass plane.}
\scriptsize
\begin{tabular}{|c||c|c|}
\hline
$\beta$-search & \multicolumn{2}{c|}{Trapezoid parameters, opening angle $\theta$=22 degrees}  \\
Target signal mass [$\GeV$]& Lower edge mass [$\GeV$] & Upper edge mass [$\GeV$]\\
\hline
\hline
150 & 120 & 210 \\
200 & 160 & 290 \\
250 & 210 & 380 \\
300 & 250 & 490 \\
350 & 270 & 640 \\
400 & 320 & 680 \\
450 & 370 & 700 \\
500 & 400 & 810 \\
550 & 470 & 930 \\
600 & 480 & 1360 \\
650 & 530 & 1360 \\
700 & 530 & 1390 \\
800 & 570 & 2380 \\
900 & 620 & 5340 \\
1000 & 720 & 6170 \\
1100 & 780 & 6170 \\
1200 & 860 & 6170 \\
1300 & 860 & 6170 \\
1400 & 950 & 7000 \\
1600 & 950 & 7000 \\
1800 & 950 & 7000 \\
2000 & 1100 & 7000 \\
2200 & 1200 & 7000 \\
2400 & 1200 & 7000 \\
2600 & 1660 & 7000 \\

\hline
\end{tabular}
\label{tab:h_trapezoid_mass_windows}
\end{table}

\FloatBarrier

\subsubsection{Di-track search mass windows}

Two masses, $\massAdEdx$ and $\massBdEdx$, can be calculated for the
pair of tracks surviving the two-track signal selection using
$\bgdEdx$ and the momentum of each track. Pair-produced sleptons should
have the same mass, and are therefore more likely than the background
to populate a trapezoid in the [$\massAdEdx$, $\massBdEdx$] plane. The
trapezoid window definition is optimised for the slepton mass region. Two cases are
considered separately: the Discovery SR, with only two mass windows created to
maximise the signal significance (see Table~\ref{tab:ditrackDiscovery_mass_windows}) and the Exclusion SR,
with mass windows optimised for the strongest signal exclusion (see
Table~\ref{tab:ditrackExclusion_mass_windows}). There is no upper mass limit to the trapezoid in
either the Discovery or Exclusion SR, as adding one was not found to
improve sensitivity in the two-track analysis. The optimal angle of
the trapezoid is found to increase slowly with target mass, as the
momentum measurement dominates the reconstructed mass at moderate mass
and above, and the reconstructed momentum of the tracks are
independent (unlike in the
$\beta$-SR.)

\begin{table}[tp]
\centering
\caption{The 2D mass window parameters for the two-track Discovery SR derived using the trapezoidal
method. The angle identifies the region of compatibility between the mass measurements, and the lower edge mass identifies the mass point
($\massAdEdx$, $\massBdEdx$) at the centre of the trapezoidal lower edge.}
\scriptsize
\begin{tabular}{|c|c|c|}
\hline
Di-track search & & \\
Target signal mass [\GeV] & Opening angle [Degrees] & Lower edge mass [\GeV]\\
\hline
$< 350$ & 45 & 160 \\
$\ge 350$ & 45 & 300 \\
\hline
\end{tabular}
\label{tab:ditrackDiscovery_mass_windows}
\end{table}

\begin{table}[tp]
\centering
\caption{The 2D mass window parameters for the two-track Exclusion SR derived using the trapezoidal method. The angle identifies the region of compatibility between the mass measurements, and the lower edge mass identifies the mass point
($\massAdEdx$, $\massBdEdx$) at the centre of the trapezoidal lower edge.}
\scriptsize
\begin{tabular}{|c|c|c|}
\hline
Di-track search & & \\
Target signal mass [\GeV] & Opening angle [Degrees] & Lower edge mass [\GeV] \\
\hline
200 & 20 & 160 \\
300 & 20 & 260 \\
400 & 26 & 340 \\
500 & 28 & 430 \\
600 & 28 & 520\\
700 & 28 & 600 \\
\hline
\end{tabular}
\label{tab:ditrackExclusion_mass_windows}
\end{table}

\FloatBarrier


\subsection{Uncertainties}
\label{subsec:systematics}

Systematic uncertainties come from several sources: the data-driven background determination, corrections for the detector effects, and the experimental and theoretical signal modelling uncertainties.

The data-driven background estimate is based on a pseudo-data method, where the $p_T$, $\eta$, \dEdx and $\betacalo$ variables are generated for each pseudo-data event. This generation method comes with the assumption that the \dEdx and $p_T$ are uncorrelated while the $\eta$ correlation is taken into account through the $\eta$-slicing of the samples. To evaluate the validity of this assumption, a closure test is implemented for the $\beta$-search where all kinematic, \dEdx, and \betacalo templates are extracted from the $\beta\gamma$-CR of the SR to generate a background distribution and compared with data in a subset of this same region after applying the same \dEdx and \betacalo cuts used for the SR. Any observed non-closure might signify that there are correlations that are not taken into account, and a \emph{template correlation uncertainty} must be assigned according to the size of this non-closure in each mass trapezoid.
As an additional check the same closure test is repeated also in the $\beta\gamma$-CR of each VR. This uncertainty ranges from 5\% at 200~\GeV to 23\% at 700~\GeV and is the largest single uncertainty in most mass windows.

For the di-track search, the dominant systematic uncertainty in the background estimate is also obtained from a closure test. The validation regions test the assumptions of the background estimation method, but the comparison between prediction and data is limited by the data statistics in the validation regions. The mass distributions from each of the validation regions is compared between the pseudo-data prediction and the observed data and used to construct a ratio; a 0.68 confidence internal band is fit across all regions from the deviation from unity in this ratio. This uncertainty is calculated and applied per mass bin; it increases with the target mass from about 10\% in the lowest mass windows to about 40\% in the highest mass window.

The choice of the $\eta$ binning has a direct effect on the background template shape, and a different $\eta$ slicing choice could impact the final results. For this reason, an alternative $\eta$ slicing\footnote{The edges of the baseline $\eta$-binning are [0, 0.1, 0.2, 0.3, ..., 1.3, 1.4, 1.5, 1.6] while the edges of the alternative $\eta$-binning are [0, 0.15, 0.25, 0.35, ..., 1.35, 1.45, 1.6].} is used to generate an alternative data-driven background and then calculate an \emph{alternative $\eta$-slice uncertainty}, which is below $5\%$ in all mass windows. Another concern with the \dEdx template is the lack of data at large \dEdx values, which can result in the large \dEdx values to be over or under represented in the data-driven background estimate. To circumvent this issue, alternative \dEdx templates are generated by fitting the original \dEdx templates with a Crystal Ball function~\cite{SUSY-2018-42} that models the \dEdx template with high accuracy for small \dEdx values and does not suffer from statistical uncertainty issues at large \dEdx values. The difference between these templates for large \dEdx values allows to calculate a \emph{\dEdx tail uncertainty}, which is everywhere below $1\%$. For the di-track analysis, these two uncertainties are found to be negligible.

The data-driven background mass distributions in the signal regions are simulated using pseudo-data events extracted from \dEdx, $\betacalo$ and kinematics templates. The statistical precision of the pseudo-data experiments is limited by the statistical uncertainty of the input templates. To compute a correct statistical uncertainty of the background templates, the statistical uncertainties of \dEdx, $\betacalo$ and kinematic templates need to be propagated to the final background distributions that are generated by pseudo-data events. To calculate the statistical uncertainty due to the background templates for the $\beta$-search, the input kinematic templates are first smoothed to remove empty bins, then these templates are randomly fluctuated assuming Poisson distributions. The newly generated templates are then used to throw new pseudo-data generating alternative background mass distributions. The root mean square difference between the alternative mass distributions is then used as the \emph{statistical uncertainty}, with a maximum value of $5\%$. For the $\beta$-search, an additional percent-level \emph{ normalisation uncertainty} is computed by propagating the statistical uncertainties of the generated background and data (which dominates this uncertainty) through the normalisation method and combining them quadratically. For the di-track search, the \dEdx template uncertainties are handled similarly, but the kinematic CR statistical uncertainty is different because the template consists of unbinned events. For the statistical uncertainty related to the finite number of events in the kinematic CR, each event is Poisson fluctuated from a nominal expectation value of one event in the sample; i.e. some events do not appear in a fluctuated sample while others appear multiple times. This is known as a Bootstrap procedure.

An \emph{\MET trigger uncertainty} in the background estimate is evaluated by re-generating the background without reweighing the events in the $\beta\gamma$-CR and comparing with the same events reweighed with an \MET trigger threshold weight. The \emph{\MET trigger uncertainty} is below 2\%. There is no \MET reweighing of the data for the di-track search, and as \MET does not play a role in the di-track background estimate, there is no associated uncertainty.

For the $\beta$-search, two systematic uncertainties are considered that evaluate the effect on the background estimate of choices made in treating the \betacalo variable. The \emph{\betacalo $\eta$ slicing uncertainty} evaluates the effect of the choice of $\eta$-binning of the \betacalo templates, and is calculated similarly to the \dEdx $\eta$ slicing uncertainty described above. The \emph{\betacalo $\eta$ slicing uncertainty} amounts to roughly 10\% in all mass windows. The other \betacalo uncertainty is similar to the \dEdx tail uncertainty. Also in this case it is necessary to evaluate  how well the background estimate models the low tail of the \betacalo distribution. This tail is not well-populated in the $\eta$-sliced \betacalo templates, and so not all $\eta$-dependent effects might be captured in the background estimate. To account for this, a new background estimate is created where \betacalo is drawn from a Crystal Ball function that is fitted to the data rather than from the low tails of the \betacalo templates. The difference between this new background estimate and the nominal background estimate defines the \emph{\betacalo tail uncertainty}, which ranges from 3\% to 10\%.

The signal cross-section uncertainties are displayed in the final limit plots as theoretical uncertainties in the excluded cross-section.
Signals with long lifetimes that are detector stable and signal models with a small mass splitting between the invisible decay product and the LLP parent mass mostly rely on the presence of an ISR jet to satisfy the online \MET trigger. For these signal models, the ISR modelling is expected to be the largest signal systematic uncertainty.
To estimate this uncertainty, alternative generator-level signal MC samples were generated with factorisation, renormalisation, and merging scales varied from the nominal values separately by 0.5 and 2.0, as well as with five variations concerning parton shower tuning and radiation uncertainty~\cite{ATL-PHYS-PUB-2014-021}. The $\pt$ of the sparticle and signal-mass dependent weights are then extracted from these samples to parameterise the differences between the samples with scale variations.
The newly acquired weights are then applied to fully reconstructed signal MC samples. The differences between the reweighed MC samples and the nominal samples is used as a systematic uncertainty. The majority of these uncertainties are small; the leading ISR jet uncertainty is around $10\%$.

Additional uncertainties in the signal selection acceptance and efficiency associated with the simulation modelling related to the modelling of the pile-up distribution, the calculation of \MET, and track-level quantities are assessed; no individual uncertainty due to the modelling of these quantities has an impact larger than a few percent on the signal yield. The modelling of the \MET trigger efficiency in simulation is validated and an uncertainty derived by comparing the trigger efficiency, as a function of \MET, between data and simulation for a sample of $Z\rightarrow\mu\mu$ events with similar track-level requirements as the tracks in the search. The difference between the observed and simulated trigger efficiency measured in $Z\rightarrow\mu\mu$ events is used to correct the signal trigger efficiency based on the reconstructed offline \MET, neglecting both muons and the soft track term. The difference between this shifted value and the nominal yield in simulation is used as an additional uncertainty in the signal yield, which is about 15\% or lower for all stau signals.

The uncertainty in the combined 2015--2018 integrated luminosity is 0.83\%~\cite{DAPR-2021-01}, obtained using the LUCID-2 detector~\cite{LUCID2} for the primary luminosity measurements, complemented by measurements using the inner detector and calorimeters.

\FloatBarrier


%
\section{Results}
\label{sec:result}

Results are presented separately for the $\beta$-search and the two-track search.
The statistical analysis and likelihood construction were implemented in the \texttt{pyhf} software framework~\cite{pyhf}. For each trapezoidal mass window, the likelihood of the background-only hypothesis given the observed data was constructed from the background prediction and the associated systematic uncertainties. The effect of the systematic uncertainties is incorporated through nuisance parameters that are constrained to be Gaussian-distributed. Using a profile-likelihood-based test statistic~\cite{Cowan:2010js}, independent $p_{0}$-values quantifying the level of agreement between the observed data and the background prediction were calculated for each of these windows.

\subsection{$\beta$-search results}
\begin{figure}[htbp]
\begin{center}
\includegraphics[width=0.60\textwidth]{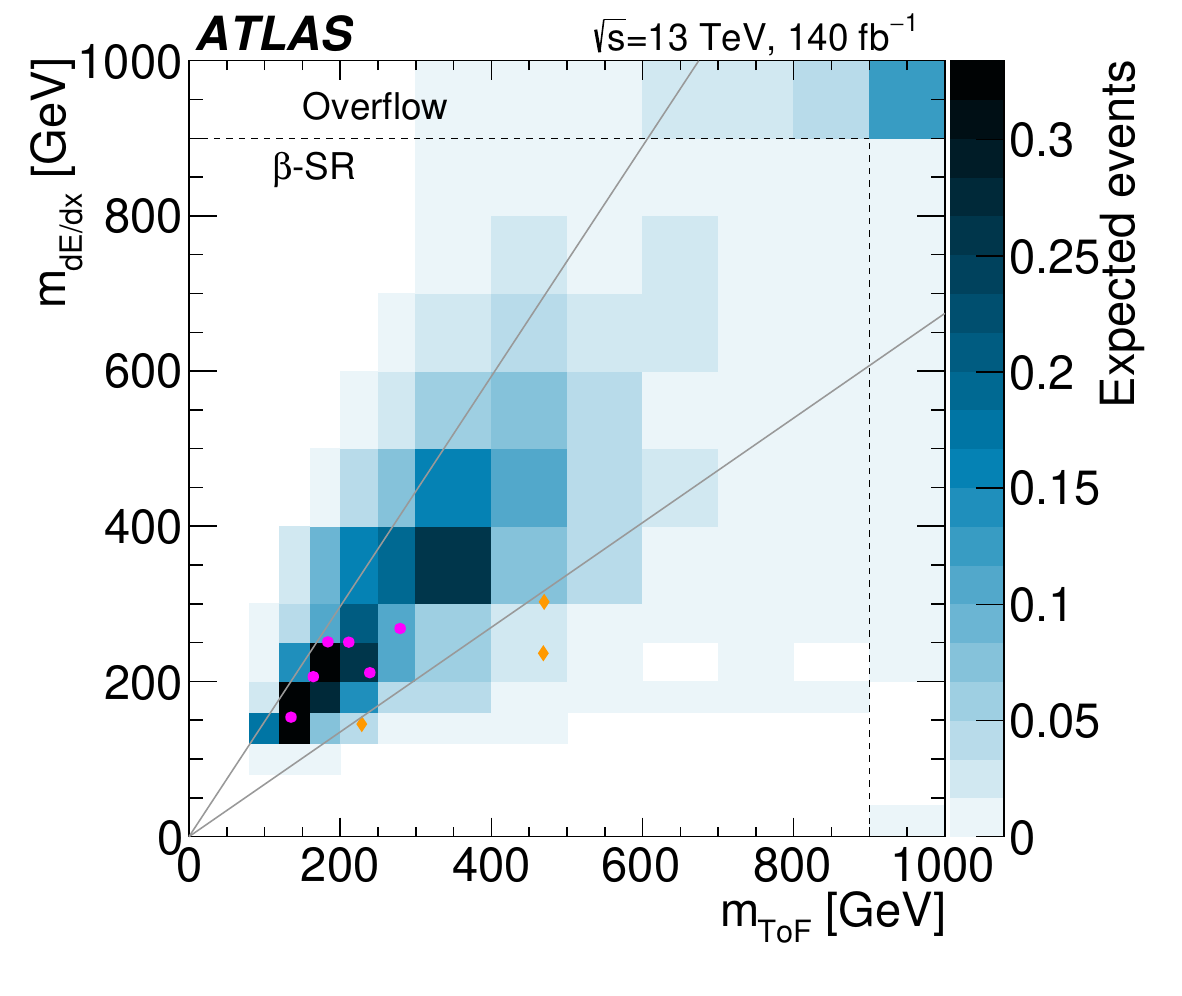}
\caption{The distribution of data and predicted background in the $\beta$-search signal region. The observed data events are indicated as magenta circles if they are inside the mass-compatibility angle (shown as grey lines) and as orange diamonds if they are outside, while the blue area is the mass distribution of the background. The overflow is included in the $900<\massdEdx<1000~\GeV$ and in the $900<\massToF<1000~\GeV$ regions.}
\label{fig:SR_BCalo_2D}
\end{center}
\end{figure}

Without any mass-compatibility requirement, nine events are observed in the signal region while the background expectation is of 5.1$\pm0.5$ events. The mass values of those events are shown as circles in the [$\massdEdx, \massToF$] plane in Figure~\ref{fig:SR_BCalo_2D}.

\begin{figure}[htbp]
\begin{center}
\includegraphics[width=0.60\textwidth]{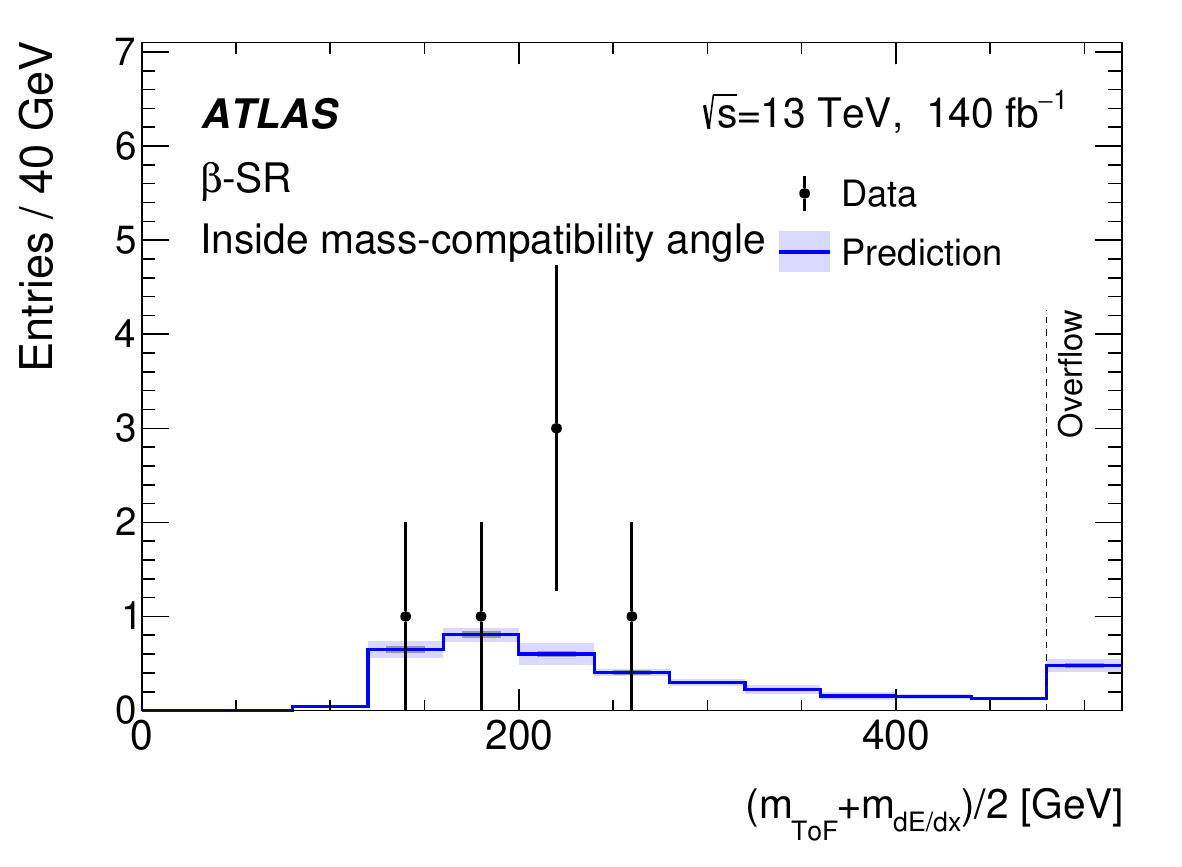}
\caption{The distribution of the average of $\massToF$ and $\massdEdx$ compared with the expected background in the $\beta$-search signal region. The systematic uncertainty in the predicted background is calculated as indicated in Section~\ref{subsec:systematics}. Only events inside the 22 degrees mass-compatibility angle are considered.}
\label{fig:average_mass_inclusive}
\end{center}
\end{figure}

Restricting the observation to be in the 22 degree mass-compatibility angle, the observed events are reduced to six and the background expectation to 3.7$\pm0.4$. The observed data and the expected background in each mass window are shown in Table~\ref{tab:BCalo_results}. The distribution of the average of $\massdEdx$ and $\massToF$ is shown in Figure~\ref{fig:average_mass_inclusive} for the events inside the mass-compatibility angle.

\begin{table}[htbp]
\centering
\caption{Data and background yields in the trapezoids defined for different masses in the $\beta$-driven analysis. The table extends up to 400~\GeV covering the mass region where there are data entries and beyond, including an overflow bin. The regions are not orthogonal.}
\scriptsize
\label{tab:BCalo_results}
\begin{tabular}{|c|c|c|} \hline
Target signal mass [\GeV] & Expected background & Observed data\\
\hline
150 & 1.73 $\pm$ 0.17 & 2 \\
200 & 1.89 $\pm$ 0.20 & 5 \\
250 & 1.40 $\pm$ 0.17 & 4 \\
300 & 1.24 $\pm$ 0.17 & 1 \\
350 & 1.23 $\pm$ 0.18 & 0 \\
400 & 0.88 $\pm$ 0.14 & 0 \\
> 400 & 0.80 $\pm$ 0.12 & 0 \\
\hline
\end{tabular}
\end{table}

The lowest p-value of $3.3\times10^{-2}$ is measured in the $200~\GeV$ mass window.

\subsection{Di-track search results}

\begin{figure}[htbp]
\begin{center}
\includegraphics[width=0.6\textwidth]{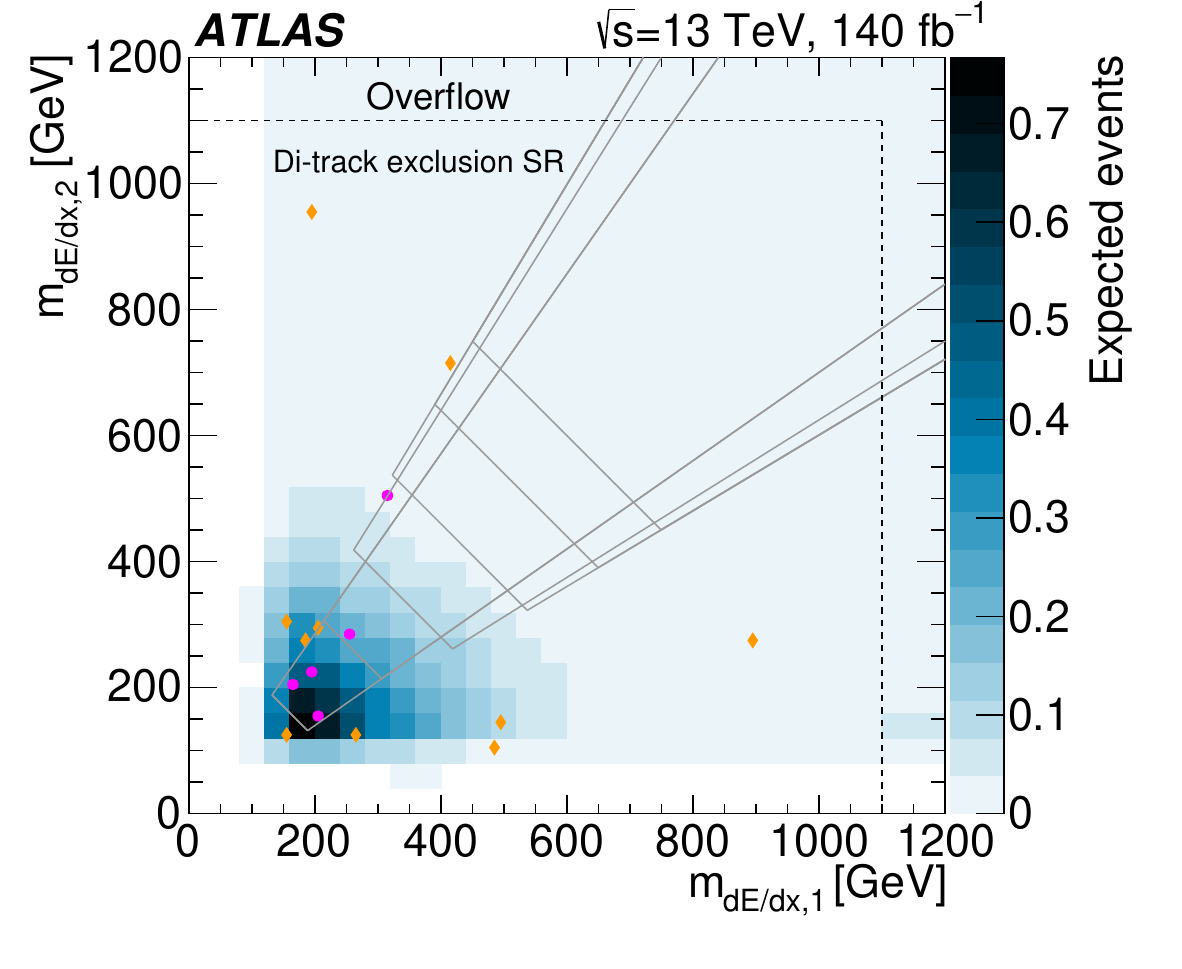}
\caption{The distribution of data and predicted background in the di-track Exclusion-SR. The observed data events are indicated as magenta circles if they are inside the mass-compatibility angle (shown as grey lines) and as orange diamonds if they are outside, while the blue area is the mass distribution of the expected background. The last bins include overflow events}
\label{fig:SR_2track_2D}
\end{center}
\end{figure}

\begin{figure}[htbp]
\begin{center}
\includegraphics[width=0.6\textwidth]{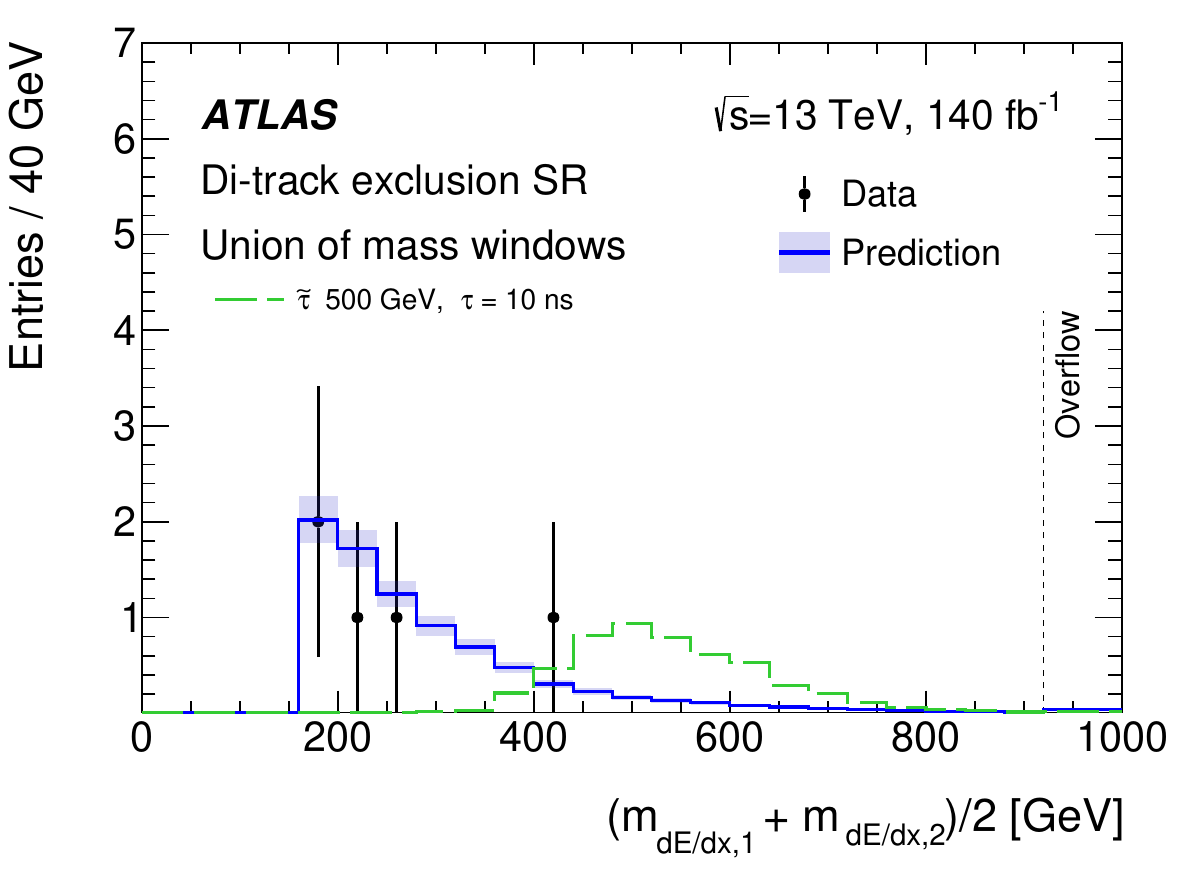}
\caption{The distribution of the average $\massdEdx$ of both tracks in each event in the union of all the mass windows in the di-track Exclusion-SR, compared with the expected background. The systematic uncertainty in the predicted background is calculated as indicated in Section~\ref{subsec:systematics}. The yield for a 500~\GeV stau signal with a lifetime of 10~ns is also shown.}
\label{fig:SR_2track_1D}
\end{center}
\end{figure}

There are 15 events observed in the inclusive Exclusion-SR region, with a background expectation of 20.7 $\pm 4.5$ events. The distribution of $\massAdEdx$ and $\massBdEdx$ for each event in the inclusive Exclusion-SR is shown in Figure~\ref{fig:SR_2track_2D}. There are five observed events that fall into the union of all mass windows; the distribution of the average $\massdEdx$ of both tracks in these 5 events is shown in Figure~\ref{fig:SR_2track_1D}. The observed data and expected background in each mass window in the Exclusion-SR are shown in Table~\ref{tab:2track_results}.

\begin{table}[tp]
\centering
\caption{The observed data and expected background in each mass window of the di-track Exclusion-SR. Both systematic and statistical uncertainties in the expected background are included. The regions are not orthogonal.}
\scriptsize
\begin{tabular}{|c|c|c|}
\hline
Target signal mass [\GeV] & Expected background & Observed data \\
\hline
200 & 7.93 $\pm$ 1.56  & 4 \\
300 & 3.49 $\pm$ 0.89  & 1 \\
400 & 2.09 $\pm$ 0.74  & 1 \\
500 & 1.07 $\pm$ 0.49  & 0 \\
600 & 0.59 $\pm$ 0.32  & 0 \\
700 & 0.35 $\pm$ 0.20  & 0 \\
\hline
\end{tabular}
\label{tab:2track_results}
\end{table}

There are zero events observed in the Discovery-SR, with an inclusive background expectation of $0.79 \pm 0.19$ events. The expected background in both Discovery-SR mass windows is shown in Table~\ref{tab:2track_DR}, along with observed and expected model-independent 95\% confidence level (CL) upper limits on the number of signal events and the observed 95\% CL upper limit on the visible cross-section. Also shown is the discovery p-value, which measures the compatibility of the observed data with the background-only hypothesis relative to fluctuations of the background.

\begin{table}[tp]
\centering
\caption{The observed data and expected background in both mass windows of the di-track Discovery-SR. Both systematic and statistical uncertainties in the expected background are included. The regions are not orthogonal. Also shown are model-independent observed and expected 95\% CL upper limits on the number of signal events, $S_{\texttt{obs}}^\texttt{95}$ and $S_{\texttt{exp}}^\texttt{95}$, the observed 95\% CL upper limit on the visible cross-section, $\sigma_{\texttt{vis}}$, and the discovery p-value, $P(s=0)$. The p-value is capped at 0.5.}
\scriptsize
\begin{tabular}{|c|c|c|c|c|c|c|}
\hline
Lower mass edge [\GeV] & Expected background & Observed data & $S_{\texttt{obs}}^\texttt{95}$ & $S_{\texttt{exp}}^\texttt{95}$ & $\sigma_{\texttt{vis}}$ [fb] & $P(s=0)$ \\
\hline
160 & 0.67 $\pm 0.17$  & 0 & 3.0 & 3.1 & 0.02 & 0.5 \\
300 & 0.29 $\pm 0.10$  & 0 & 3.0 & 3.0 & 0.02 & 0.5 \\
\hline
\end{tabular}
\label{tab:2track_DR}
\end{table}

\subsection{Lifetime-dependent mass limits}

The results of this study are interpreted for the benchmark signal models considered, and the 95\% CL upper limit on the cross-section is extracted using pseudo-experiments and the $\text{CL}_\text{s}$ prescription~\cite{Read:2002hq} for each signal mass and lifetime hypothesis. Cross-section limits as a function of lifetime, for varying signal masses, and cross-sections limits as a function of signal mass, for varying lifetimes, are provided for all benchmark signal models in Ref.~\cite{HEPdata}.

To more accurately probe the sensitivity of the analysis to LLP lifetimes other than those used in the generation of the signal samples, the same samples are reinterpreted for intermediate lifetime values by reweighing the LLP particle decay spectra. Intermediate lifetimes are modelled by reweighing the closest longer-lifetime sample to shorter lifetimes, except for $\tau>30~\text{ns}$. The choice of target lifetimes for $\tau>30~\text{ns}$ is limited by the reduced size of the reweighed sample.

The di-track search is optimised for the stau scenario. It has higher sensitivity than the $\beta$-search in this domain, but lower sensitivity for charginos and $R$-hadrons. The mass limits reported below are therefore obtained with the di-track analysis for staus and with the $\beta$-driven analysis for charginos and $R$-hadrons.

\begin{figure}[tbp]
\begin{center}
\subfigure[Stau]{
\label{fig:stau_limits}
\includegraphics[width=0.47\textwidth]{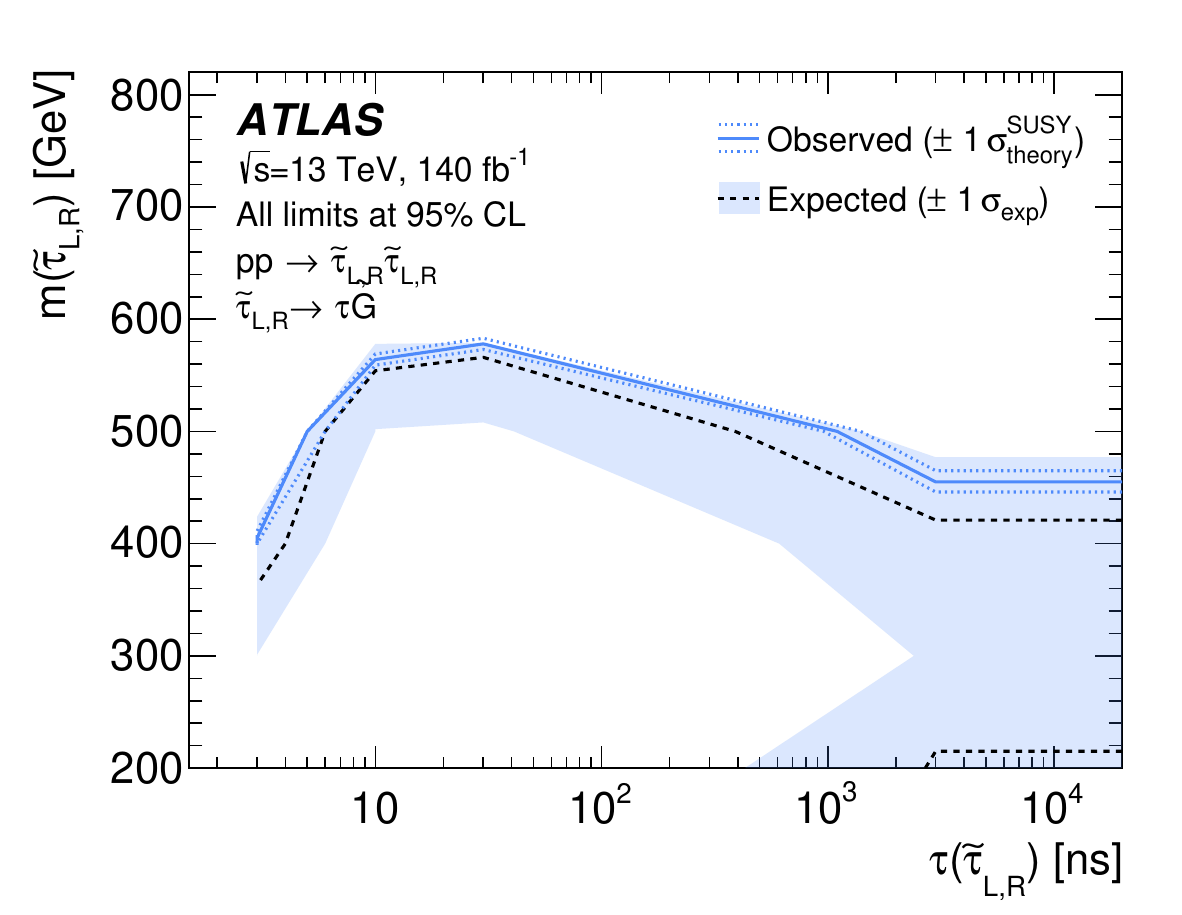}
}
\subfigure[Chargino]{
\label{fig:chargino_2Dlimit_hadd}
\includegraphics[width=0.47\textwidth]{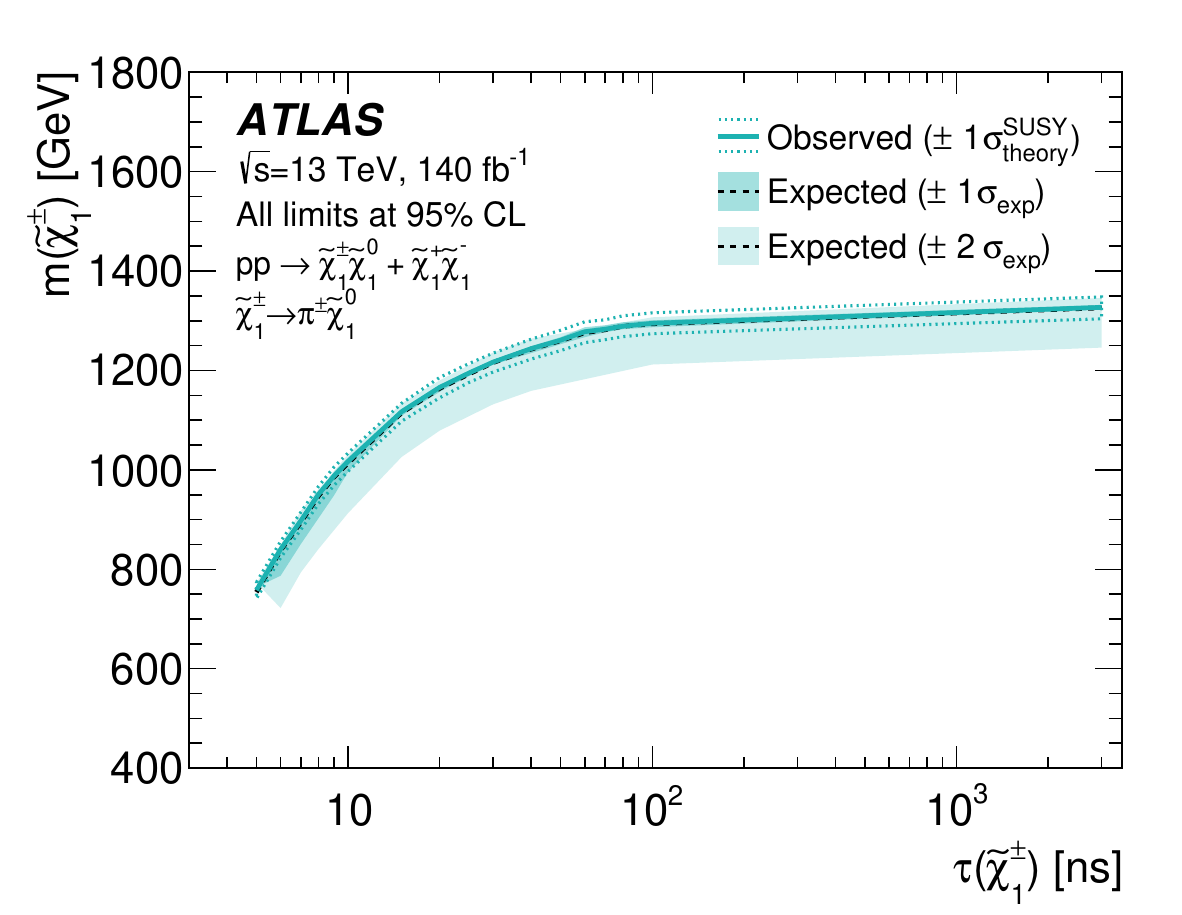}
}
\caption{%
(a) The contour for the excluded mass--lifetime region for stau pair production obtained with the di-track search. All masses and lifetimes shown that are below the curve and above 200~\GeV are excluded by the observed data (while the expected exclusion is between the upper curve down to 210~\GeV for lifetimes above 3000~ns). The sensitivity extends indefinitely to longer lifetimes. (b) Upper limits on the chargino mass, assuming the $\chinoonepm\ninoone$ and $\chinoonep\chinoonem$ contributions, versus lifetime obtained with the $\beta$-search. Lifetimes below 3~ns and masses below 200~\GeV are not probed by these analyses. Observed limits are indicated as solid blue lines and expected limits are indicated by dotted black lines. The shaded band around the expected limit indicate the 1$\sigma_{\text{exp}}$ (and 2$\sigma_{\text{exp}}$) uncertainty range, derived as explained in Section~\ref{subsec:systematics}.}
\end{center}
\end{figure}

The limits in the stau scenario are shown in Figure~\ref{fig:stau_limits}. While the interpretation is done only for staus, the analysis has similar sensitivity for selectrons and smuons. Differences between signal selection efficiency arise only from second-order effects on \met and isolation criteria from the different interactions of the slepton decay products with the detector. The sensitivity for staus peaks at around 30~ns for two reasons: at lower lifetimes, the LLPs do not travel far enough to be reconstructed as tracks, while at higher lifetimes, the \MET trigger efficiency drops, as discussed in Section~\ref{subsec:selection}.

The mass range $200\mbox{--}560~\GeV$ is excluded for mass-degenerate $\stau_\mathrm{L}$ and $\stau_\mathrm{R}$ with lifetimes $\tau=10~\text{ns}$, while the corresponding expected exclusion is $200\mbox{--}550~\GeV$. This search is not sensitive to masses below 200~\GeV as lighter LLPs have lower \pt and lower $\dEdx$ which do not allow for reasonable background discrimination. These results are the most sensitive to date for detector-unstable $\stau_\mathrm{L,R}$ with lifetimes above $3~\text{ns}$. At lower lifetimes, the ATLAS search for displaced leptons provides exclusions for $\stau_\mathrm{L,R}$ with $\tau =0.3~\text{ns}$ up to 380~\GeV~\cite{SUSY-2022-11}. Searches for detector-stable LLPs, exploiting the muon system as a trigger, have previously been performed, including an ATLAS search for detector-stable LLPs with $\dEdx$ and ToF using $36$~fb$^{-1}$ of data that excluded stable nearly pure $\stau_\mathrm{R}$ with masses up to 430~\GeV~\cite{SUSY-2016-32}, and a CMS result using 101~fb$^{-1}$ of data that excluded detector-stable $\stau_\mathrm{R}$ up to 520~\GeV and $\stau_\mathrm{L,R}$ with masses up to 730~\GeV~\cite{CMSLLP2024}.

Figure~\ref{fig:chargino_2Dlimit_hadd} shows the mass limits for sum of the $\chinoonepm\ninoone$ and $\chinoonep\chinoonem$ production. The chargino mass limit is $\approx1.3~\TeV$ for lifetimes $\tau>100~\text{ns}$. These results provide the most stringent limits to date for detector-unstable charginos in the lifetime range above $10~\text{ns}$. The previous ATLAS search that selected a single track with significant \dEdx had greater sensitivity for charginos with lifetimes from $3~\text{ns}$ to $10~\text{ns}$~\cite{SUSY-2018-42}, as the requirement for the LLPs to travel to the hadronic calorimeter is less efficient for signals with shorter lifetimes.

\begin{figure}[tbp]
\begin{center}
\subfigure[]{
\label{fig:rhadron-limits-normal}
\includegraphics[width=0.47\textwidth]{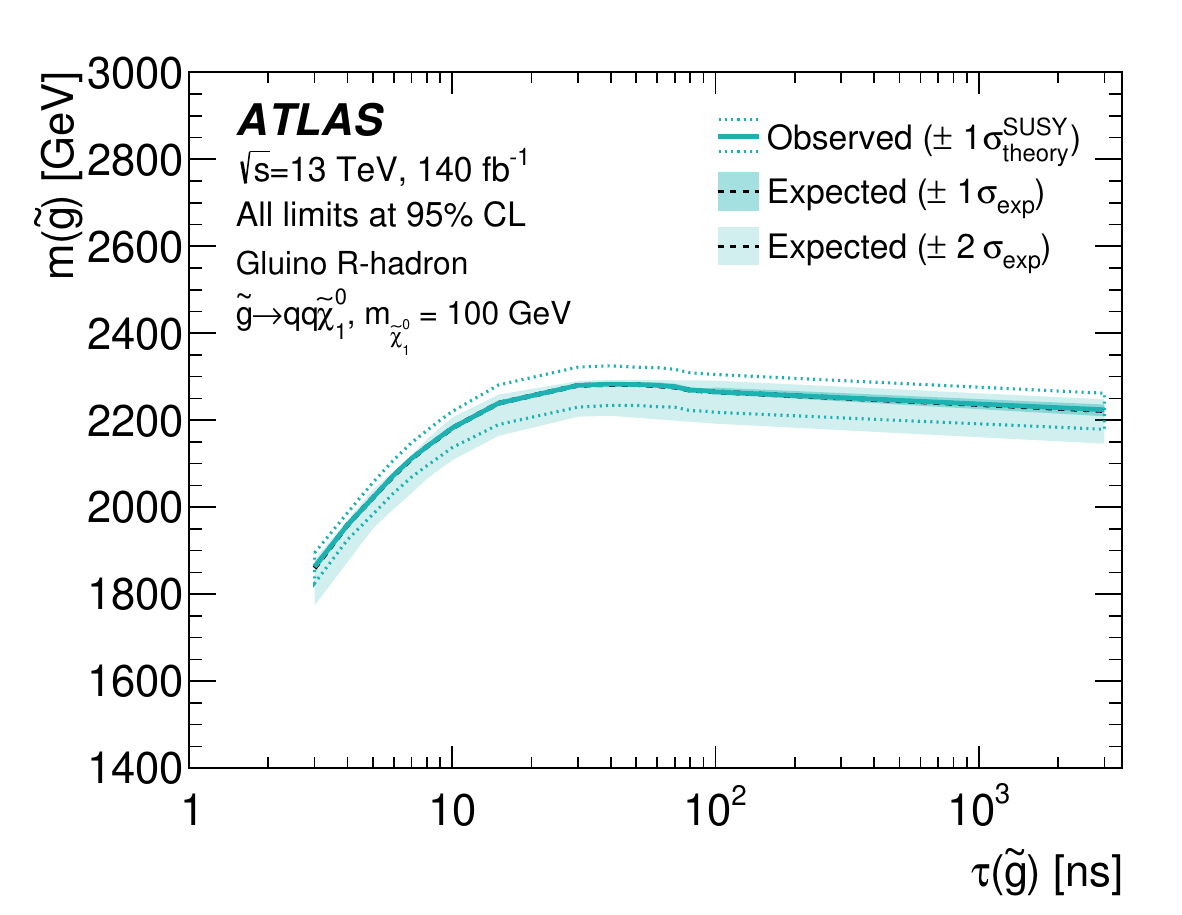}
}
\subfigure[]{
\label{fig:rhadron-limits-compressed}
\includegraphics[width=0.47\textwidth]{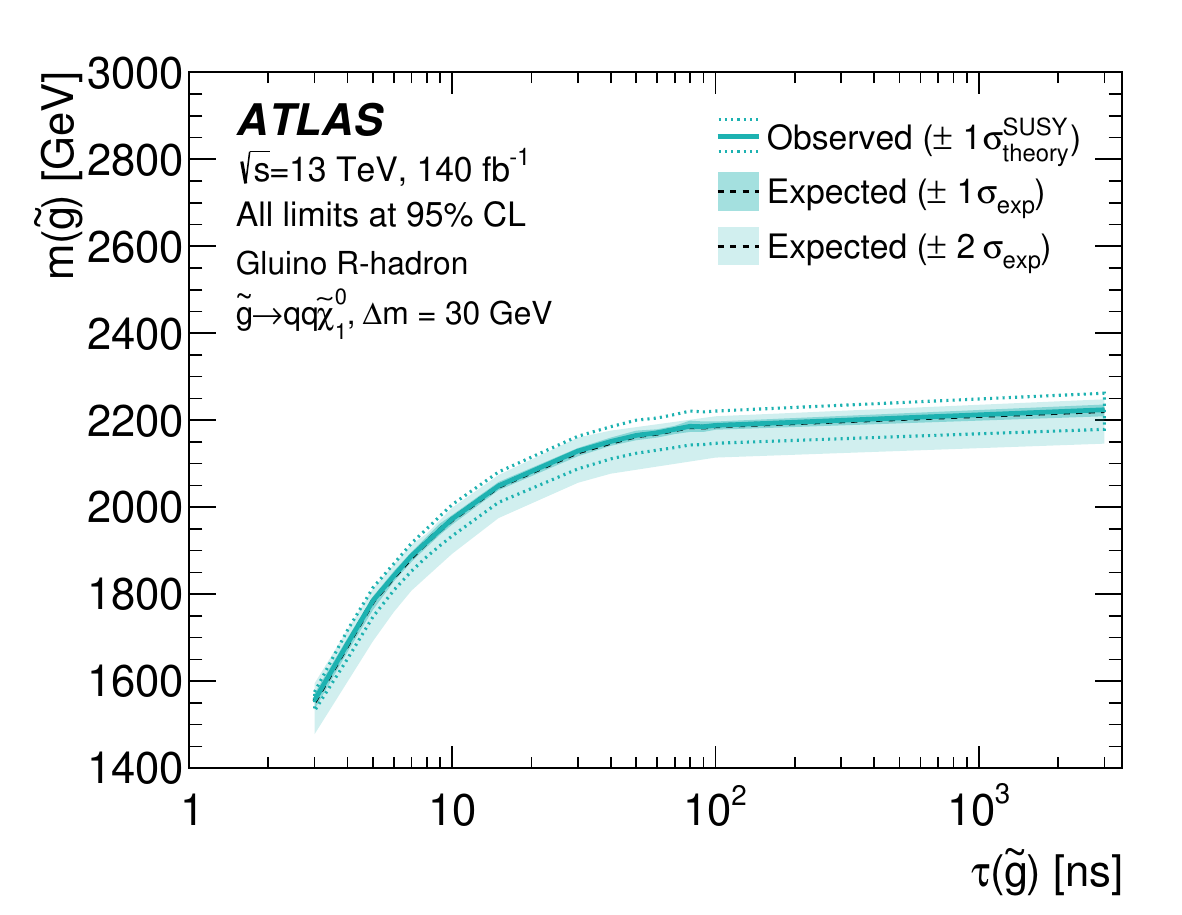}
}
\caption{Upper limits on the gluino mass, from gluino $R$-hadron pair production, as a function of the gluino lifetime for two neutralino mass assumptions of \subref{fig:rhadron-limits-normal} $m(\ninoone)=100~\GeV$ and \subref{fig:rhadron-limits-compressed} $\Delta m(\gluino,\ninoone)=30~\GeV$ obtained with the $\beta$-search. Observed limits are indicated as solid blue lines and expected limits are indicated by dotted black lines. The shaded band around the expected limit indicate the 1$\sigma_{\text{exp}}$ (and 2$\sigma_{\text{exp}}$) uncertainty range, derived as explained in Section~\ref{subsec:systematics}. The upper 1$\sigma_\text{exp}$ expected bound is very close to the expected limit for some lifetime values due to the expected background approaching zero events. For a given lifetime, the mass values below the curve are excluded.}
\label{fig:Fake_RH}
\end{center}
\end{figure}

Figure~\ref{fig:Fake_RH} shows the mass limits for gluino $R$-hadron pair production for both the $m(\ninoone) = 100~\GeV$ and $\Delta m(\gluino,\ninoone)=30~\GeV$ cases. The sensitivity for $R$-hadrons with $m(\ninoone) = 100~\GeV$ falls for lifetimes above 30~ns due to a loss of efficiency for the \MET trigger, as discussed in Section~\ref{subsec:selection}. The charginos and $R$-hadrons with $\Delta m(\gluino,\ninoone)=30~\GeV$ do not have decay products that interact significantly with the detector, so their \MET trigger efficiency is flat as a function of lifetime. The highest observed lower limit on the mass is $2.27~\TeV$ ($2.20~\TeV$) and is obtained at $\tau=30~\text{ns}$ ($\tau>200~\text{ns}$) for $m(\ninoone) = 100~\GeV$ $\left(\Delta m(\gluino,\ninoone)=30~\GeV\right)$, while the corresponding expected limit matches the observed limit. These results provide the most stringent limits to date for detector-unstable LLPs in the lifetime range above $10~\text{ns}$. At lifetimes lower than $10~\text{ns}$, other strategies including searches for disappearing tracks~\cite{CMS-SUS-21-006,SUSY-2018-19} and searches for displaced vertices~\cite{CMS-EXO-22-020,SUSY-2018-13} provide better sensitivity.

\FloatBarrier


%

%
%
\FloatBarrier

\section{Conclusion}
\label{sec:conclusion}

A search is performed for heavy charged LLPs of lifetime exceeding $3~\text{ns}$ produced at the LHC in $140~\mathrm{fb^{-1}}$ of $pp$ collisions at $\sqrt{s}=13~\TeV$.
This search is based on two independent and compatible measurements of the LLP mass compared to a data-driven background estimate. The two mass measurements are obtained either for two heavily-ionising opposite-sign particles or for one heavily ionising particle that is also measured to be slow-moving. The mass values are determined by the $\beta\gamma$ measurements obtained either through the specific ionisation in the pixel detector or through the ToF measured by the hadronic calorimeter.
Two independent determinations of $\beta\gamma$ minimise the effect of the fluctuations that happen in the far tails of those distributions.

Observed yields and distributions agree with the SM background expectations and limits are placed on several simplified SUSY models.
The highest sensitivity is reached for LLPs with lifetimes exceeding 10~$\text{ns}$. Masses smaller than $2.27~\TeV$ are excluded at the 95\% confidence level for gluino $R$-hadrons with a lifetime of 30~$\text{ns}$ and $m(\ninoone)=100~\GeV$. The mass limit for compressed-scenario $R$-hadrons, with $\Delta m(\gluino,\ninoone)=30~\GeV$ and a lifetime $>200~\text{ns}$, is $2.20~\TeV$. Charginos with masses smaller than 1.3~$\TeV$ and lifetime $>100~\text{ns}$ are excluded. Masses in the range of $200\mbox{--}560~\GeV$ for staus are excluded for lifetimes of 10~$\text{ns}$.

The limits for detector unstable LLPs in the mass--lifetime plane are the most stringent to date in the lifetime domain exceeding 10~$\text{ns}$ and provide further constraints on the $R$-hadron, chargino and stau production models considered.

The seven events in the 3.3  Z significance excess observed in the signal region defined by Ref.~\cite{SUSY-2018-42} are excluded by the $\betacalo$ selection. This indicates that this excess is not due to heavy, highly-ionising and slow particles reaching the hadronic calorimeter.

\FloatBarrier


\FloatBarrier
\section*{Acknowledgements}
%

%

%
%

%
%

We thank CERN for the very successful operation of the LHC and its injectors, as well as the support staff at
CERN and at our institutions worldwide without whom ATLAS could not be operated efficiently.

The crucial computing support from all WLCG partners is acknowledged gratefully, in particular from CERN, the ATLAS Tier-1 facilities at TRIUMF/SFU (Canada), NDGF (Denmark, Norway, Sweden), CC-IN2P3 (France), KIT/GridKA (Germany), INFN-CNAF (Italy), NL-T1 (Netherlands), PIC (Spain), RAL (UK) and BNL (USA), the Tier-2 facilities worldwide and large non-WLCG resource providers. Major contributors of computing resources are listed in Ref.~\cite{ATL-SOFT-PUB-2025-001}.

We gratefully acknowledge the support of ANPCyT, Argentina; YerPhI, Armenia; ARC, Australia; BMWFW and FWF, Austria; ANAS, Azerbaijan; CNPq and FAPESP, Brazil; NSERC, NRC and CFI, Canada; CERN; ANID, Chile; CAS, MOST and NSFC, China; Minciencias, Colombia; MEYS CR, Czech Republic; DNRF and DNSRC, Denmark; IN2P3-CNRS and CEA-DRF/IRFU, France; SRNSFG, Georgia; BMBF, HGF and MPG, Germany; GSRI, Greece; RGC and Hong Kong SAR, China; ICHEP and Academy of Sciences and Humanities, Israel; INFN, Italy; MEXT and JSPS, Japan; CNRST, Morocco; NWO, Netherlands; RCN, Norway; MNiSW, Poland; FCT, Portugal; MNE/IFA, Romania; MSTDI, Serbia; MSSR, Slovakia; ARIS and MVZI, Slovenia; DSI/NRF, South Africa; MICIU/AEI, Spain; SRC and Wallenberg Foundation, Sweden; SERI, SNSF and Cantons of Bern and Geneva, Switzerland; NSTC, Taipei; TENMAK, T\"urkiye; STFC/UKRI, United Kingdom; DOE and NSF, United States of America.

Individual groups and members have received support from BCKDF, CANARIE, CRC and DRAC, Canada; CERN-CZ, FORTE and PRIMUS, Czech Republic; COST, ERC, ERDF, Horizon 2020, ICSC-NextGenerationEU and Marie Sk{\l}odowska-Curie Actions, European Union; Investissements d'Avenir Labex, Investissements d'Avenir Idex and ANR, France; DFG and AvH Foundation, Germany; Herakleitos, Thales and Aristeia programmes co-financed by EU-ESF and the Greek NSRF, Greece; BSF-NSF and MINERVA, Israel; NCN and NAWA, Poland; La Caixa Banking Foundation, CERCA Programme Generalitat de Catalunya and PROMETEO and GenT Programmes Generalitat Valenciana, Spain; G\"{o}ran Gustafssons Stiftelse, Sweden; The Royal Society and Leverhulme Trust, United Kingdom.

In addition, individual members wish to acknowledge support from Armenia: Yerevan Physics Institute (FAPERJ); CERN: European Organization for Nuclear Research (CERN DOCT); Chile: Agencia Nacional de Investigaci\'on y Desarrollo (FONDECYT 1230812, FONDECYT 1230987, FONDECYT 1240864); China: Chinese Ministry of Science and Technology (MOST-2023YFA1605700, MOST-2023YFA1609300), National Natural Science Foundation of China (NSFC - 12175119, NSFC 12275265, NSFC-12075060); Czech Republic: Czech Science Foundation (GACR - 24-11373S), Ministry of Education Youth and Sports (ERC-CZ-LL2327, FORTE CZ.02.01.01/00/22\_008/0004632), PRIMUS Research Programme (PRIMUS/21/SCI/017); EU: H2020 European Research Council (ERC - 101002463); European Union: European Research Council (ERC - 948254, ERC 101089007, ERC, BARD, 101116429), Horizon 2020 Framework Programme (MUCCA - CHIST-ERA-19-XAI-00), European Union, Future Artificial Intelligence Research (FAIR-NextGenerationEU PE00000013), Italian Center for High Performance Computing, Big Data and Quantum Computing (ICSC, NextGenerationEU); France: Agence Nationale de la Recherche (ANR-20-CE31-0013, ANR-21-CE31-0013, ANR-21-CE31-0022, ANR-22-EDIR-0002); Germany: Baden-Württemberg Stiftung (BW Stiftung-Postdoc Eliteprogramme), Deutsche Forschungsgemeinschaft (DFG - 469666862, DFG - CR 312/5-2); Italy: Istituto Nazionale di Fisica Nucleare (ICSC, NextGenerationEU), Ministero dell'Università e della Ricerca (NextGenEU I53D23000820006 M4C2.1.1); Japan: Japan Society for the Promotion of Science (JSPS KAKENHI JP22H01227, JSPS KAKENHI JP22H04944, JSPS KAKENHI JP22KK0227, JSPS KAKENHI JP23KK0245); Norway: Research Council of Norway (RCN-314472); Poland: Ministry of Science and Higher Education (IDUB AGH, POB8, D4 no 9722), Polish National Agency for Academic Exchange (PPN/PPO/2020/1/00002/U/00001), Polish National Science Centre (NCN 2021/42/E/ST2/00350, NCN OPUS 2023/51/B/ST2/02507, NCN OPUS nr 2022/47/B/ST2/03059, NCN UMO-2019/34/E/ST2/00393, UMO-2020/37/B/ST2/01043, UMO-2021/40/C/ST2/00187, UMO-2022/47/O/ST2/00148, UMO-2023/49/B/ST2/04085, UMO-2023/51/B/ST2/00920); Portugal: Foundation for Science and Technology (FCT); Spain: Generalitat Valenciana (Artemisa, FEDER, IDIFEDER/2018/048), Ministry of Science and Innovation (MCIN \& NextGenEU PCI2022-135018-2, MICIN \& FEDER PID2021-125273NB, RYC2019-028510-I, RYC2020-030254-I, RYC2021-031273-I, RYC2022-038164-I); Sweden: Carl Trygger Foundation (Carl Trygger Foundation CTS 22:2312), Swedish Research Council (Swedish Research Council 2023-04654, VR 2018-00482, VR 2021-03651, VR 2022-03845, VR 2022-04683, VR 2023-03403), Knut and Alice Wallenberg Foundation (KAW 2018.0458, KAW 2019.0447, KAW 2022.0358); Switzerland: Swiss National Science Foundation (SNSF - PCEFP2\_194658); United Kingdom: Leverhulme Trust (Leverhulme Trust RPG-2020-004), Royal Society (NIF-R1-231091); United States of America: U.S. Department of Energy (ECA DE-AC02-76SF00515), Neubauer Family Foundation.

%
%


%

\clearpage
\printbibliography
\clearpage
 
\begin{flushleft}
\hypersetup{urlcolor=black}
{\Large The ATLAS Collaboration}

\bigskip

\AtlasOrcid[0000-0002-6665-4934]{G.~Aad}$^\textrm{\scriptsize 105}$,
\AtlasOrcid[0000-0001-7616-1554]{E.~Aakvaag}$^\textrm{\scriptsize 17}$,
\AtlasOrcid[0000-0002-5888-2734]{B.~Abbott}$^\textrm{\scriptsize 124}$,
\AtlasOrcid[0000-0002-0287-5869]{S.~Abdelhameed}$^\textrm{\scriptsize 120a}$,
\AtlasOrcid[0000-0002-1002-1652]{K.~Abeling}$^\textrm{\scriptsize 56}$,
\AtlasOrcid[0000-0001-5763-2760]{N.J.~Abicht}$^\textrm{\scriptsize 50}$,
\AtlasOrcid[0000-0002-8496-9294]{S.H.~Abidi}$^\textrm{\scriptsize 30}$,
\AtlasOrcid[0009-0003-6578-220X]{M.~Aboelela}$^\textrm{\scriptsize 46}$,
\AtlasOrcid[0000-0002-9987-2292]{A.~Aboulhorma}$^\textrm{\scriptsize 36e}$,
\AtlasOrcid[0000-0001-5329-6640]{H.~Abramowicz}$^\textrm{\scriptsize 158}$
% ...
\bigskip
\\

% ...
$^{17}$Department for Physics and Technology, University of Bergen, Bergen; Norway.\\
% ...
$^{30}$Physics Department, Brookhaven National Laboratory, Upton NY; United States of America.\\
% ...
$^{36}$$^{(a)}$Facult\'e des Sciences Ain Chock, Universit\'e Hassan II de Casablanca;$^{(b)}$Facult\'{e} des Sciences, Universit\'{e} Ibn-Tofail, K\'{e}nitra;$^{(c)}$Facult\'e des Sciences Semlalia, Universit\'e Cadi Ayyad, LPHEA-Marrakech;$^{(d)}$LPMR, Facult\'e des Sciences, Universit\'e Mohamed Premier, Oujda;$^{(e)}$Facult\'e des sciences, Universit\'e Mohammed V, Rabat;$^{(f)}$Institute of Applied Physics, Mohammed VI Polytechnic University, Ben Guerir; Morocco.\\
% ...
$^{46}$Physics Department, Southern Methodist University, Dallas TX; United States of America.\\
% ...
$^{50}$Fakult\"{a}t Physik , Technische Universit{\"a}t Dortmund, Dortmund; Germany.\\
% ...
$^{56}$II. Physikalisches Institut, Georg-August-Universit\"{a}t G\"ottingen, G\"ottingen; Germany.\\
% ...
$^{105}$CPPM, Aix-Marseille Universit\'e, CNRS/IN2P3, Marseille; France.\\
% ...
$^{120}$$^{(a)}$New York University Abu Dhabi, Abu Dhabi;$^{(b)}$United Arab Emirates University, Al Ain; United Arab Emirates.\\
% ...
$^{124}$Homer L. Dodge Department of Physics and Astronomy, University of Oklahoma, Norman OK; United States of America.\\
% ...
$^{158}$Raymond and Beverly Sackler School of Physics and Astronomy, Tel Aviv University, Tel Aviv; Israel.\\
% ...
$^{a}$ Also Affiliated with an institute covered by a cooperation agreement with CERN.\\
$^{b}$ Also at An-Najah National University, Nablus; Palestine.\\
$^{c}$ Also at Borough of Manhattan Community College, City University of New York, New York NY; United States of America.\\
$^{d}$ Also at Center for High Energy Physics, Peking University; China.\\
$^{e}$ Also at Center for Interdisciplinary Research and Innovation (CIRI-AUTH), Thessaloniki; Greece.\\
$^{f}$ Also at CERN, Geneva; Switzerland.\\
% ...

\end{flushleft}


%

%
%
%
\FloatBarrier
\clearpage

\end{document}